\documentclass[a4paper,11pt]{article}
\usepackage{amsfonts}

\pdfoutput=1

\usepackage{graphicx}
\usepackage{amsmath}
\usepackage{amssymb}
\usepackage{latexsym}
\usepackage[colorlinks]{hyperref}
\usepackage{color}
\usepackage{float}
\usepackage{cite}
\usepackage{makeidx}
\usepackage{colortbl}
\usepackage{csquotes}
\usepackage[squaren]{SIunits}

\usepackage[textfont={footnotesize,sf},labelfont={color=blue,bf,sf},labelsep=endash]{caption}
\usepackage[position=top,labelfont={color=blue,bf,sf}]{subfig}
\usepackage{bm}

\newcommand{\bra}{\begin{array}}
\newcommand{\era}{\end{array}}
\newcommand{\beq}{\begin{equation}}
\newcommand{\eeq}{\end{equation}}
\newcommand{\bqr}{\begin{eqnarray}}
\newcommand{\eqr}{\end{eqnarray}}

\def\BC{\bb C}
\def\_\BC{\bbi C}

\def\Tr {{\rm Tr}}

\def\( {\left(}
\def\) {\right)}
\def\no2 {{\textstyle{n\over 2}}}

\def\Tr {{\rm Tr}}

\begin{document}

\begin{titlepage}
\setcounter{page}{1}
\renewcommand{\thefootnote}{\fnsymbol{footnote}}

\begin{flushright}
\end{flushright}

\vspace{5mm}
\begin{center}

{\Large \bf {Transport Properties in Graphene Superlattices}}

\vspace{5mm}
{\bf El Bou\^azzaoui Choubabi}$^{a}$, {\bf Abdellatif Kamal}$^{a}$ and
 {\bf Ahmed Jellal\footnote{\sf 
a.jellal@ucd.ac.ma}}$^{a,b}$

\vspace{5mm}

{$^{a}$\em Laboratory of Theoretical Physics,  
Faculty of Sciences, Choua\"ib Doukkali University},\\
{\em PO Box 20, 24000 El Jadida, Morocco}


{$^b$\em Saudi Center for Theoretical Physics, Dhahran, Saudi Arabia}

\vspace{30mm}

\begin{abstract}

            Using Chebyshev polynomials, we study the electronic transport properties of massless
            Dirac fermions in symmetrical graphene superlattice composed of three regions. Matching 
            wavefunctions and using transfer matrix method, we explicitly determine transmission probability 
            as well as the conductance and Fano factor. At vertical Dirac points, we numerically 
            find that the transmission
            probability shows transmission gaps,  conductance has minimums and  
            Fano factor has maximums.

        \end{abstract}
        \vspace{3cm}
            
\end{center}
\noindent PACS numbers:    73.63.-b; 73.23.-b; 72.80.Rj\\
            \noindent Keywords: Graphene superlattice, transmission, conductance, Fano factor.
\end{titlepage}

\section{Introduction}

Historically, superlattices  started with semiconductor materials
and later on  have been extended to graphene systems called graphene superlattices. They 
resulted from  graphene submitted to any periodic modulation caused by
electrostatic potentials
\cite{PhysRevB.76.075430,Barbier2008, Park2008, Tiwari2009, Wang2010, PhysRevB.81.075438, Maksimova2012},
magnetic barriers modulation \cite{Sankalpa2009,Ramezani2008,Ramezani2009,DellAnna2009} and others.
Experimentally,
graphene superlattices  may be fabricated by applying a local top gate voltage 
to graphene \cite{35} or by periodically embedding impurity atoms with scanning tunneling
microscopy on graphene surface \cite{36}.
%
%
Because of their interests, recently graphene superlattices 
have motivated intense experimental and theoretical investigations 
\cite{Sankalpa2009,Barbier2008,Wang2010,Guo2011,ESMAILPOUR2010655}. 
Indeed, 
many works were devoted to study
the electronic band structures of Dirac fermions in graphene superlattices
\cite{PhysRevB.76.075430,PhysRevLett.103.046809,PhysRevB.81.075438,Arovas2010,PhysRevB.83.195434}.
%
In  graphene superlattices, it has been found that the periodic potential causes
  additional Dirac points  in the band structures and leads to an anisotropy in group
velocity of charge carriers 
causing the collimation
of electrons beams\cite{Park2008,Wang2010,PhysRevB.81.075438,Wang2011,Maksimova2012,Park2_2008,Bliokh2009}.
These results allow graphene superlattices to be a candidate for manufacturing the carbon-based nanoelectronic
devices.

Shot noise is as a consequence of the quantization of
charge and is useful to obtain information on a system, which is
not available through conductance measurements.
It is characterized by a dimensionless parameter $F$ called  Fano factor, which is 
defined as 
the ratio of noise power to mean current
\cite{47}. 
The null value $F=0$ shows a full correlation and maximal
quantum coherence giving rise to a total transmission.
When $F$ reaches a Poissonian
value $(F=1)$,
%
the transmission probability is close to zero.  
It found that 
for short and wide graphene strips 
$F$ has a maximum value
of $1/3$ at 
Dirac point,
which is similar strength to the diffusive metals \cite{Trauzettel2006}.
Recently,
we have showed that new Dirac points appeared in graphene superlattices
and their positions depend sensitively on the barrier parameters \cite{kamal2018}.
Because of such dependence, it is a good task  to 
study the  conductance
and $F$ associated to these new Dirac points.

%
%

In our previous work \cite{kamal2018},
we have studied the electronic band structures of massless Dirac fermions in 
symmetrical graphene superlattice with
cells of three regions. Using the transfer matrix method, we have  determined the dispersion relation in terms of
different physical parameters. We have numerically analyzed such relation 
and show that there exist three zones: bound,
unbound and forbidden states. In the central zone of the band structures, 
we have determined and enumerated the vertical
Dirac points $(k_y = 0)$, opening gaps and additional Dirac points. Finally, 
we have inspected the potential effect on minibands,
the anisotropy of group velocity and the energy bands contours near Dirac points. 
We have also discussed the evolution of
gap edges and cutoff region near the vertical Dirac points.


We extend our work \cite{kamal2018} to deal with other issues related to Dirac
fermions in symmetrical graphene superlattices
with cells of three regions (SSLGSL-3R). Using the transfer matrix method,
we show that the transmission amplitude can be written
in terms of the second kind Chebyshev polynomials \cite{masonchebyshev}.
After getting the current density of incident, reflected and transmitted waves along the $x$-direction,
we end up with the transmission probability, conductance and Fano factor.
Our numerical results show that at the position of vertical Dirac points 
the transmission probability has several gaps,
conductance has minimums and Fano factor has maximums. 
These 
can be controlled externally by tuning on distance $q_2$ of the central region of elementary cell,
which is the most interesting parameter of our theory. We report different discussions 
and comparison with respect to SSLGSL with two regions corresponding to $q_2=0$.

The present paper is organized as follows. In section \ref{Sec:Model}, we show how 
derive the transmission probability 
using transfer matrix method together with 
the second kind Chebyshev polynomials for Dirac fermions in 
SSLGSL-3R. Subsequently, we  numerically analyze
the transmission probability in terms of the physical parameters of our system
with an arbitrary number of cells.
%
 In section 3, we use the transmission probability to determine
 the conductance and Fano factor. 
 These quantities are obtained for a graphene superlattice with 
a number $n=30$ of elementary cells.
Our numerical
results are discussed in details to highlight the relationship between the Dirac point
locations and the transmission gaps 
as well as conductance minimums and Fano factor maximums.
We conclude our results in the final section.

\section{Transmission probability}\label{Sec:Model}

\begin{figure}[!ht]\centering
	\includegraphics[scale=0.6]{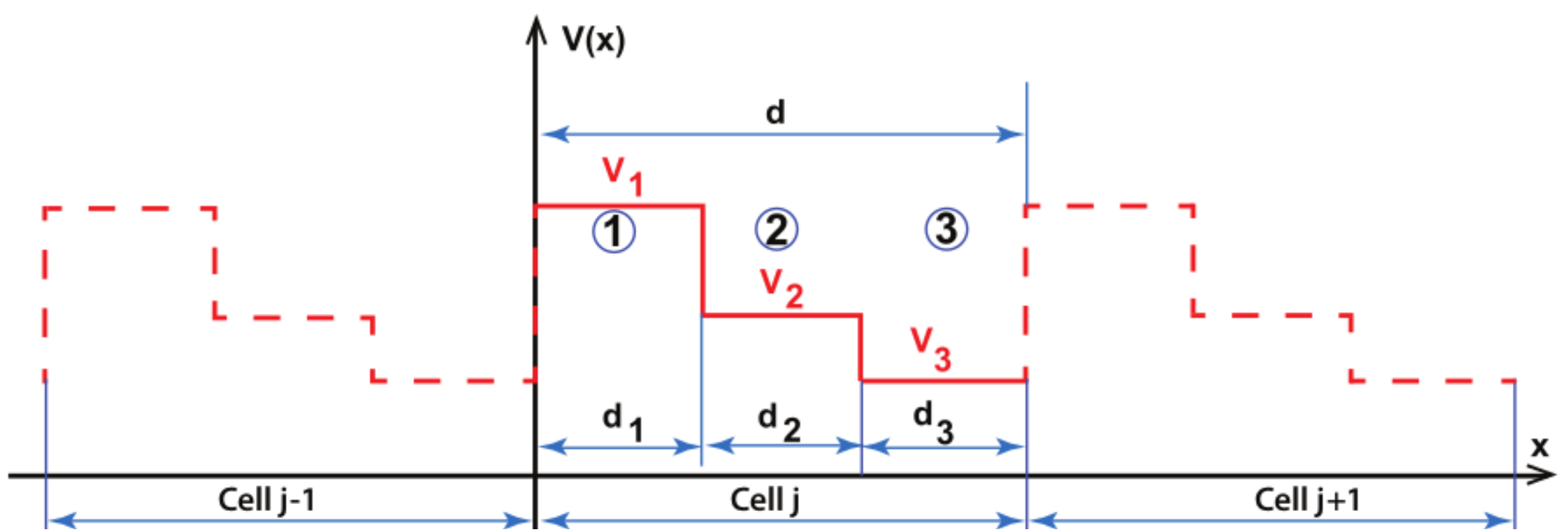}
	\caption{
		(Color online) Schematic of the superlattice potential $V(x)$ composed of three
		regions growing along the $x$-direction with the period $d=d_1+d_2+d_3$, $d_i$ is
		the width of region $i$ and $V$ is its applied potential amplitude.
	}
	\label{FigModel}
\end{figure}

We consider  massless Dirac fermions with incident energy $E$ from the \emph{input} region
of the symmetrical graphene superlattice with cells of three regions (SSLGSL-3R) at angle $\theta$, with respect
to the $x$-direction as shown in Figure \ref{FigModel}.
The periodic structure consists of $n$ elementary cell labeled by $j=0,\cdots, n-1$
where each one is composed by a juxtaposition of three single square barriers with different
height $(V_1, V_2, V_3)$ and width $(d_1,d_2 ,d_3)$, $d=d_1+d_2 +d_3$ is the width of the entire
cell. We apply a potential $V(x)$ in $j^{th }$ elementary cell (Figure \ref{FigModel})
\begin{equation}
	V^j(x)=
	\left\{
		\begin{array}{ccc}
			V_1,   & jd < x < d_1+jd\\
			V_2,   & d_1+jd < x < d_1+d_2+jd\\
			V_3,   & d_1+d_2+jd < x < (1+j)d.\\
        \end{array}
	\right.
\end{equation}

\noindent
Our graphene superlattice consists of $n$ elementary cells, which are interposed between
the \textit{input} and \textit{output} regions. The massless Dirac fermion in each region
$i$ of $j^{th }$ elementary cell is governed by the Hamiltonian
\begin{equation}
	H_{i}^{j}=v_F\bm{\sigma}\cdot\bm{p}+V_{i}^{j}\mathbb{I}
\end{equation}
where $\bm{p}=(p_x,p_y)$ is the momentum operator, $v_F\approx 10^6 \meter\per\second$ is
the Fermi velocity, $\bm{\sigma}=(\sigma_x,\sigma_y)$ are the Pauli matrices, $\mathbb{I}$
is the $2\times 2$ unit matrix, the index $i$ is running from $1$ to $3$.
The Hamiltonian acts on two components of pseudospinor
$\psi_{i}(x,y) =
	\left(%
		\begin{array}{cc}
			\varphi_i^A & \varphi_i^B \\
		\end{array}%
	\right)^{T}$
where $\varphi_i^{A/B}$ are smooth enveloping functions for two triangular sublattices ($A$, $B$)
in graphene and take the forms $\varphi_i^{A/B}(x) e^{\textbf{\emph{i}}k_y y}$ due to the translation
invariance in the $y$-direction. It is convenient to introduce the dimensionless quantities
$\mathbb{V}_i=V_i/E_F$, $\varepsilon_i=E_i/E_F$ with $E_F=\hbar v_F/d$ and therefore getting the general solution
\begin{eqnarray}
  \psi_{i}(x,y) = \psi_i(x)e^{\textbf{\emph{i}}k_y y} =
    w_i(x)D_i e^{\textbf{\emph{i}}k_y y}
\end{eqnarray}
where both matrices are
\begin{equation}\label{equ13}
	w_{i}(x)=
	\left(%
		\begin{array}{cc}
			e^{ \textbf{\emph{i}} k_{i} x} & e^{ -\textbf{\emph{i}} k_{i} x} \\
			s_{i} e^{ \textbf{\emph{i}} \theta_{i}} e^{ \textbf{\emph{i}} k_{i} x}  & -s_{i}
			e^{ -\textbf{\emph{i}} \theta_{i}} e^{- \textbf{\emph{i}} k_{i} x} \\
		\end{array}
	\right),\qquad D_i=
	\left(%
		\begin{array}{c}
			\alpha_i \\
			\beta_i \\
		\end{array}
	\right)
\end{equation}
with $s_i=\text{sign}(\varepsilon-\mathbb{V}_i)$, $\theta_{i} = \arctan \left(\frac{k_y}{k_{i}}\right)$. 
The parameters $ \alpha_i$, $\beta_i$ are the amplitude of positive and negative propagation wavefunctions
inside the region $i$, respectively.
The wave vector for region $i$ takes the form
\begin{equation}\label{h0}
  k_i=\frac{1}{d}\sqrt{(\varepsilon-\mathbb{V}_i)^2-(k_yd)^2}.
\end{equation}
Using the boundary conditions of wavefunctions at interfaces, we obtain transfer matrix 
associated with $n$ identical unit cells \cite{kamal2018}
\begin{equation}\label{T17}
	\mathcal{T}_{n}=w_{0}^{-1}(0)\Omega^{n}w_{0}(nd)
\end{equation}
where $\Omega$ reads as
\begin{equation}\label{omega}
    \Omega=w_1(0)w_1^{-1}(d_1)w_2(d_1)w_2^{-1}(d_1+d_2)w_3(d_1+d_2)w_3^{-1}(d).
\end{equation}
Calculating the determinant and trace of $\Omega$ to obtain
\begin{eqnarray}
	\det(\Omega) &=& 1\label{18}\\
	\Tr(\Omega) &=& 2\left[\cos(k_1d_1)\cos(k_2d_2)\cos(k_3d_3)+G_{12}\sin(k_1d_1)
	\sin(k_2d_2)\cos(k_3d_3)\right.\label{21}\\
    &&+\left.G_{13}\sin(k_1d_1)\sin(k_3d_3)\cos(k_2d_2)
    +G_{23}\sin(k_2d_2)\sin(k_3d_3)\cos(k_1d_1)\right]\nonumber
\end{eqnarray}
where we have introduced the quantity
\begin{equation}
	G_{ij}=\frac{(\mathbb{V}_i-\mathbb{V}_j)^2-(k_i^2+k_j^2)d^2}{2k_ik_jd^2},\qquad k_{i,j}\neq 0.
\end{equation}
With 
\eqref{21}, we can show that  the dispersion relation expression takes the form
\begin{equation}\label{Dispersion}
	\cos(k_x d)=\frac{1}{2}\Tr(\Omega).
\end{equation}


To determine the transmission probability 
it is convenient to write the transfer matrix in terms of  Chebyshev polynomials.
Then 
using \eqref{T32} (see Appendix) to show that
\eqref{T17} can be mapped as 
\begin{equation}\label{362}
	\mathcal{T}_{n}= w_0^{-1}(0)
	\left(%
		\begin{array}{cc}
			U_{n-1} \Omega_{11}-U_{n-2}  &  U_{n-1} \Omega_{12} \\
			U_{n-1} \Omega_{21} &  U_{n-1} \Omega_{22}-U_{n-2} \\
		\end{array}%
    \right)w_0(nd)
\end{equation}
which can be calculated to obtain
\begin{equation}
	\mathcal{T}_n=
	\left(%
		\begin{array}{cc}
			\mathcal{T}_{n_{11}} &\mathcal{T}_{n_{12}} \\
			\mathcal{T}_{n_{21}}  & \mathcal{T}_{n_{22}}  \\
		\end{array}%
	\right).
\end{equation}
We recall that 
the amplitudes $D_{in}=
	\left(%
		\begin{array}{c}
			1 \\
			r_n \\
		\end{array}%
    \right)$ and $ D_{out}=
    \left(%
		\begin{array}{c}
			t_n \\
			0 \\
		\end{array}%
	\right)$
of the eigenspinors in {\it input} and {\it output} regions are connected by the following  relation
\begin{equation}\label{35}
	D_{in}=\mathcal{T}_{n} D_{out}
\end{equation}
On the other hand,
the wave vector of the reflected wave along $x$-direction is opposite to that $k_{in}$ of
the incident wave and the corresponding  $\theta_{in}$ angle is transformed into
$\pi-\theta_{in}$ \cite{Jellal2012}. With this we can write the eigenspinors as
\begin{equation}
    \psi_i(x) =\alpha_i \psi^+_i(x) +\beta_i \psi^-_i(x)
\end{equation}
where 
the two components are given by
\begin{equation}
    \psi^+_i(x) =
    \left(
        \begin{array}{c}
            1 \\
            s_i e^{\textbf{\emph{i}} \theta_i}
        \end{array}
    \right)e^{i k_i x},\qquad
    \psi^-_i(x)=
    \left(
        \begin{array}{c}
            1 \\
            -s_{i} e^{ -\textbf{\emph{i}} \theta_i}
        \end{array}
    \right)
        e^{-i k_i x}.
\end{equation}
The current density corresponding to our system can be obtained as
\begin{equation} \label{J}
    \bm J=\psi^\dagger \bm\sigma \psi
\end{equation}
where $\psi$ stands for $\psi^{\sf inc}=\psi^+_{in}(x)$, $\psi^{\sf ref}= r_n~\psi^-_{in}(x)$
and $\psi^{\sf tra}= t_n~\psi^+_{out}(x)$. Using these  to derive the
incident, reflected and transmitted current density components
\begin{eqnarray} \label{J1}
J_x^{\sf inc} = 2s_{in}\cos\theta_{in}, \qquad   
 J_x^{\sf ref} = 2|r_n|^2 s_{in}\cos\theta_{in}, \qquad 
J_x^{\sf tra} = 2|t_n|^2 s_{out}\cos\theta_{out}
\end{eqnarray}
giving rise to the transmission and reflection probabilities
\begin{eqnarray} \label{J12}
T_n =\frac{\left|J_x^{\sf tra} \right|}{\left|J_{x}^{\sf inc} \right|}
=|t_n|^2, \qquad 
R_n =\frac{\left|J_{x}^{\sf ref} \right|}{\left|J_{x}^{\sf inc} \right|} =|r_n|^2
\end{eqnarray}
because 
we have ($\theta_{in}=\theta_{out}=\theta$, $k_{in}=k_{out}=k$, $s_{in}=s_{out}=s$) 
due to the symmetry of potential configuration in the \emph{input} and \emph{output} regions.
 The solution
of \eqref{35} provides the transmission coefficient $t_n$ in terms of transfer matrix element
\begin{equation}\label{361B}
	t_{n}=\frac{1}{\mathcal{T}_{n_{11}}}=\frac{e^{-i kd n}\left(1+e^{2 i\theta}\right)}{U_{n-1}-U_{n-2}-
	\Omega_{11}+2 s e^{i\theta} (U_{n-1}-\Omega_{12})-e^{2 i \theta} \Omega_{22}}
\end{equation}
showing that
 \begin{equation}\label{361}
	T_{n}=\frac{1}{|\mathcal{T}_{n_{11}}|^{2}}.
\end{equation}

\begin{figure}[!ht]\centering
	\subfloat[$n=1$]{
		\includegraphics[scale=0.42]{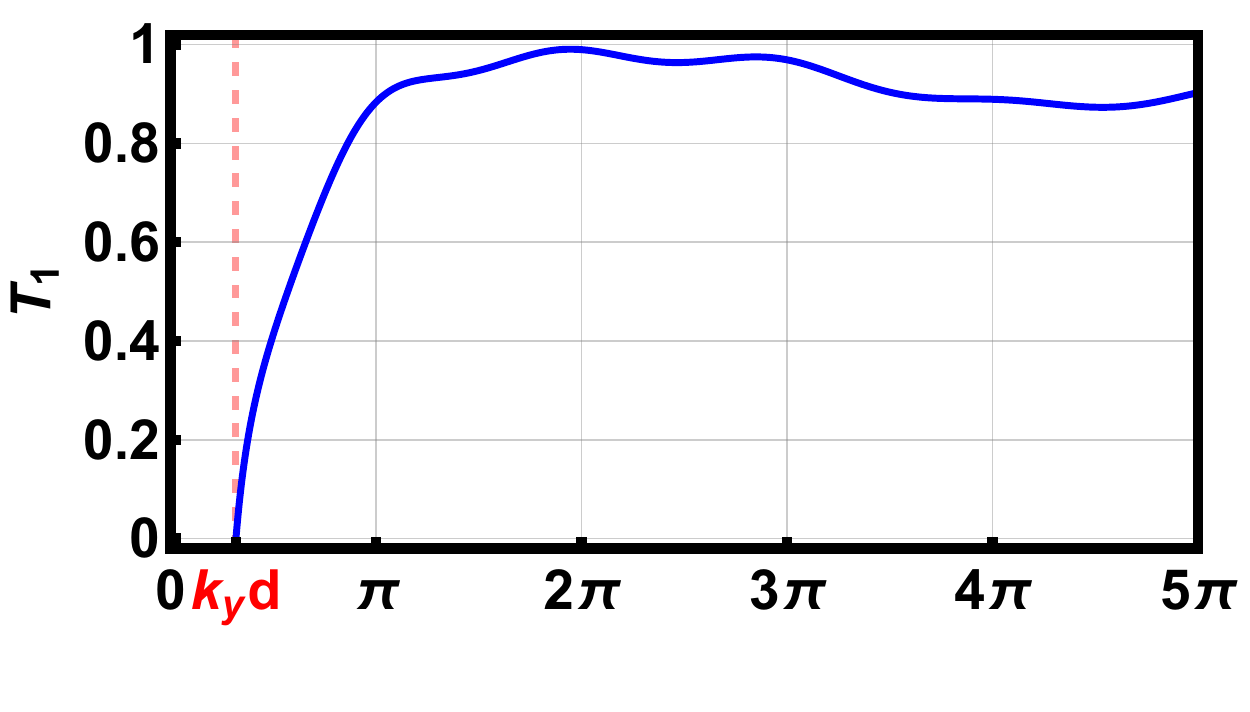}
		\label{FigTN:SubFigA}
	}\hspace{-0.45cm}
	\subfloat[$n=2$]{
		\includegraphics[scale=0.42]{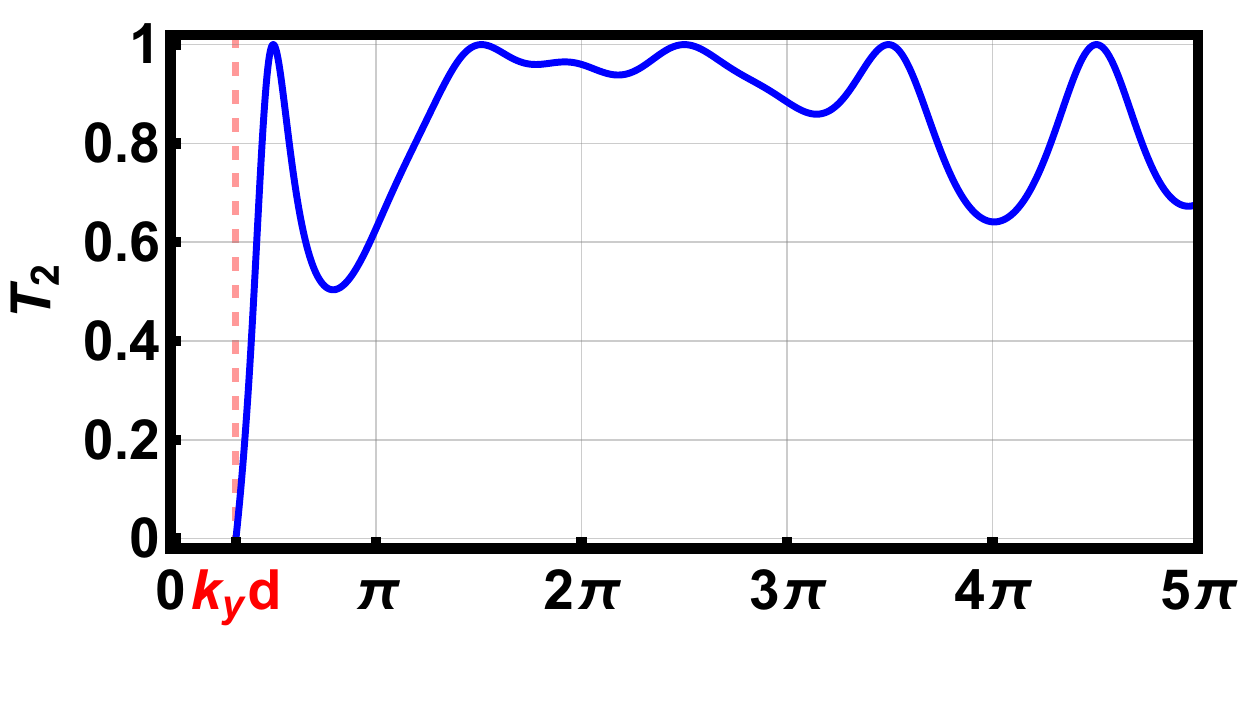}
		\label{FigTN:SubFigB}
	}\hspace{-0.45cm}
	\subfloat[$n=3$]{
		\includegraphics[scale=0.42]{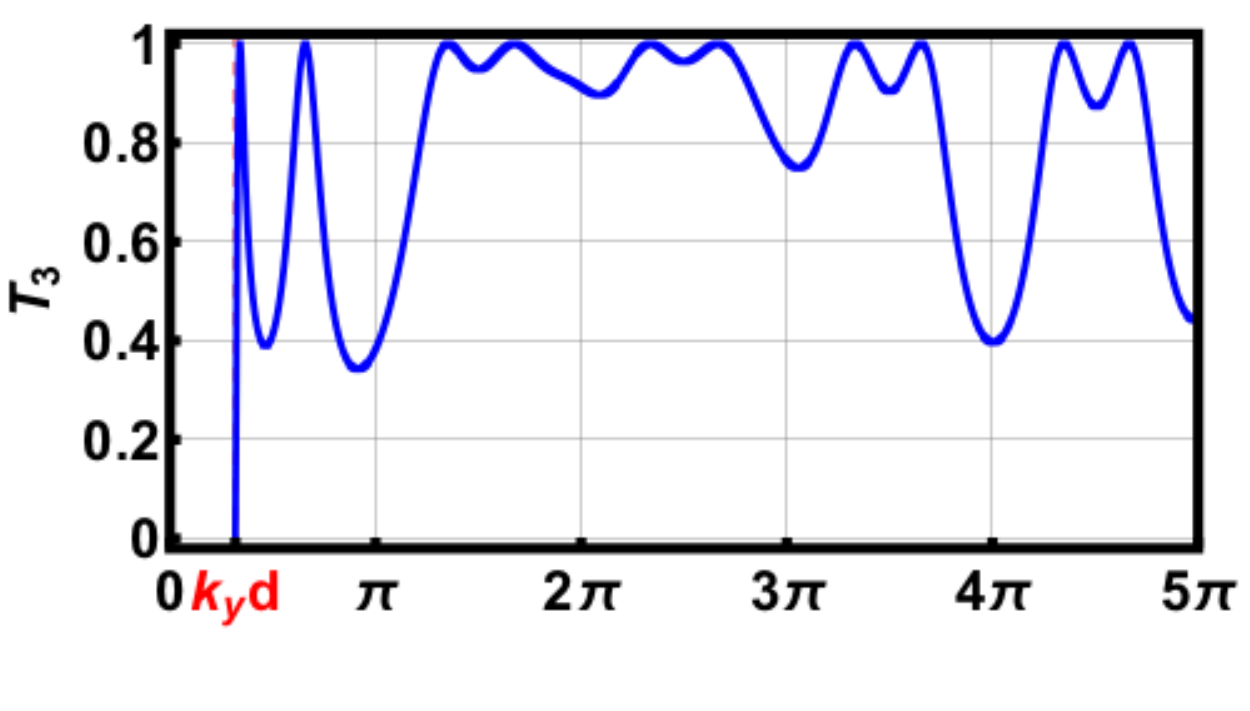}
		\label{FigTN:SubFigC}
	}\vspace{-0.7cm}
	\subfloat[$n=10$]{
		\includegraphics[scale=0.42]{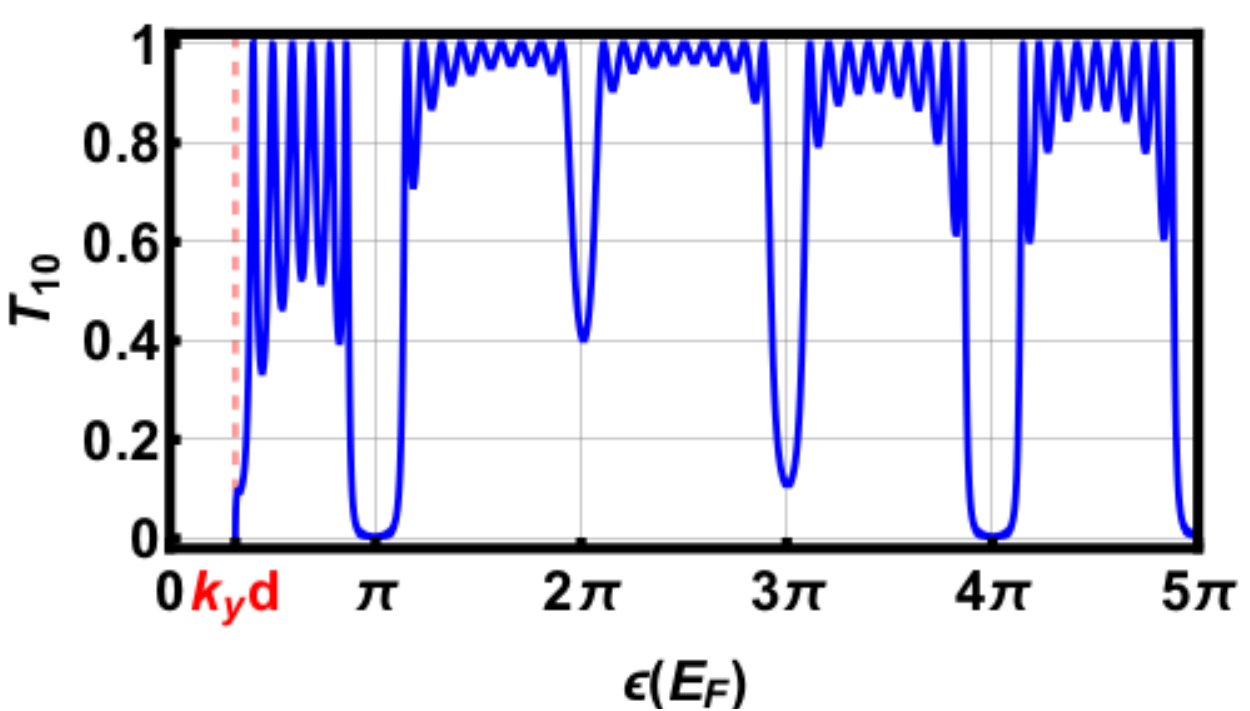}
		\label{FigTN:SubFigD}
	}\hspace{-0.45cm}
	\subfloat[$n=20$]{
		\includegraphics[scale=0.42]{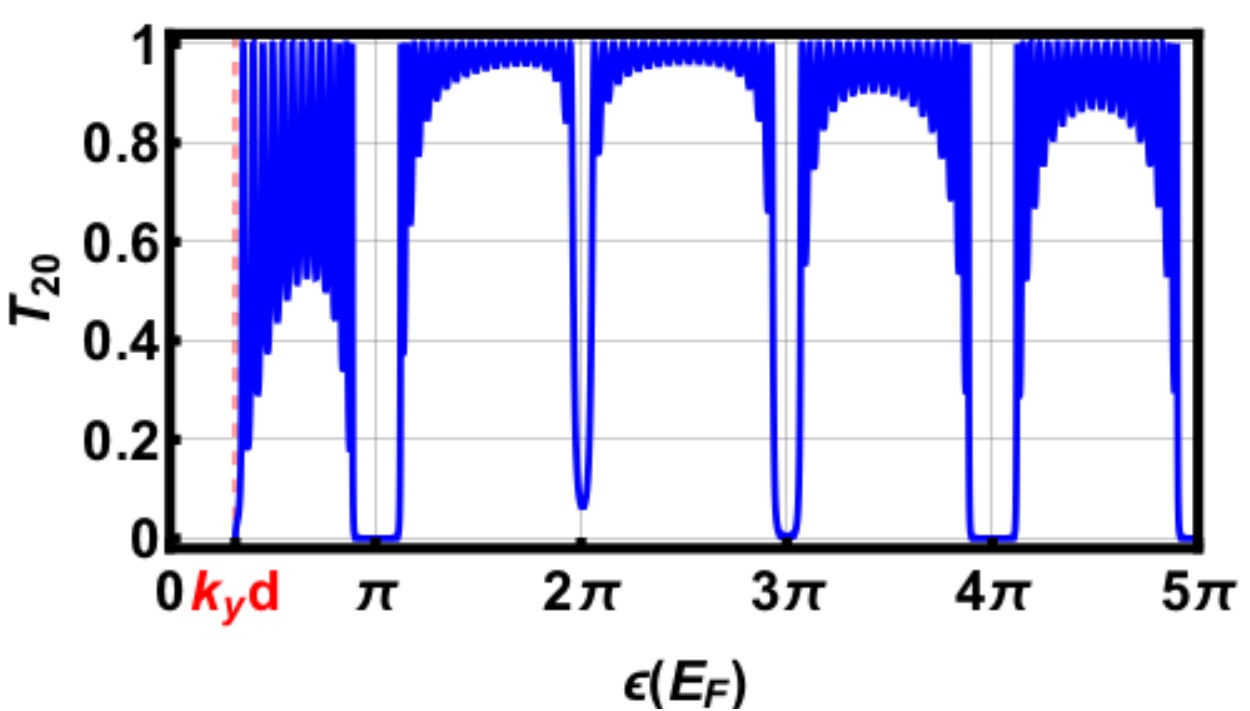}
		\label{FigTN:SubFigE}
	}\hspace{-0.45cm}
	\subfloat[$n=30$]{
		\includegraphics[scale=0.42]{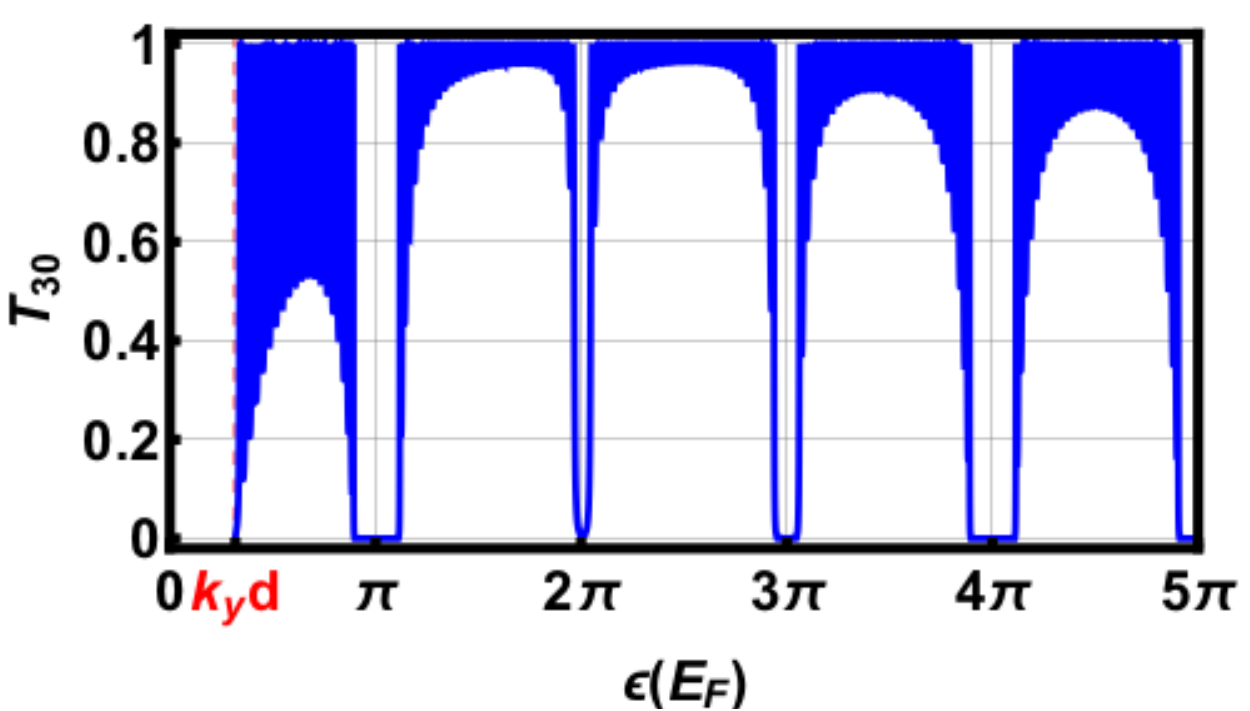}
		\label{FigTN:SubFigF}
	}
	\caption{
		(Color online) Effect of number $n$ of elementary cells  on the transmission probability
		versus incident energy $\varepsilon$  with
		$k_yd=1$, $q_2=\frac{1}{3}$, $\mathbb{V}=5\pi$.
	}
\label{FigTN}
\end{figure}

For the numerical analysis, 
we introduce the rescaled distances $0\!\leq\! q_i\!=\!\frac{d_i}{d} \!\leq\! 1$
($i=1, 2, 3$) and 
in order to carry out our computations,
we study $n$ identical elementary cells of SSLGSL-3R
with the conditions
$\left(q_1=\frac{1-q_2}{2},q_2,q_3=\frac{1-q_2}{2},
\mathbb{V}_1=-\mathbb{V}_3\equiv \mathbb{V}, \mathbb{V}_2\equiv0\right)$,
$d=10~\nano\meter$, $k_y=0.1~\nano\meter^{-1}$, $\varepsilon=m\pi$ $(m\in \mathbb{N})$.
In Figures \ref{FigTN},
we present transmission probability $T_n$ 
as function of  incident energy $\varepsilon$ for different values of number of elementary cells
$n=1,2,3,10,20,30$. We observe that $T_n$ is clearly
affected by
number $n$,
but its profile converges towards that of
superlattice when $n$ increases sufficiently up to
$n=30$. When $n$ increases, transmission gaps appear at $\varepsilon=m\pi$ 
and
become deeper, the number of oscillations
outside the transmission gaps increases and oscillations reach the total transmission. 
We notice that similar behavior has been reported in  \cite{XU2015188}.

\begin{figure}[!ht]\centering
	\subfloat[$\mathbb{V}=\pi$]{
		\includegraphics[width=5.5cm]{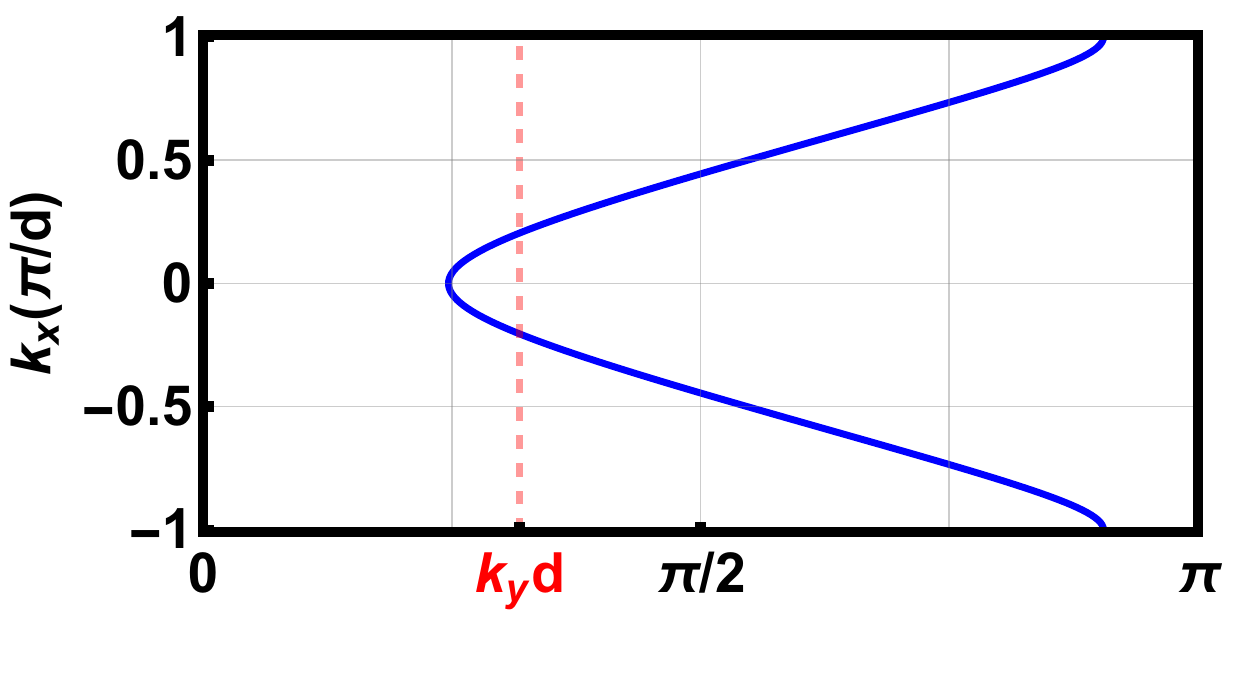}
		\label{FigKx:SubFigA}
	}\hspace{-0.51cm}
 	\subfloat[$\mathbb{V}=3\pi$]{
 		\includegraphics[width=5.5cm]{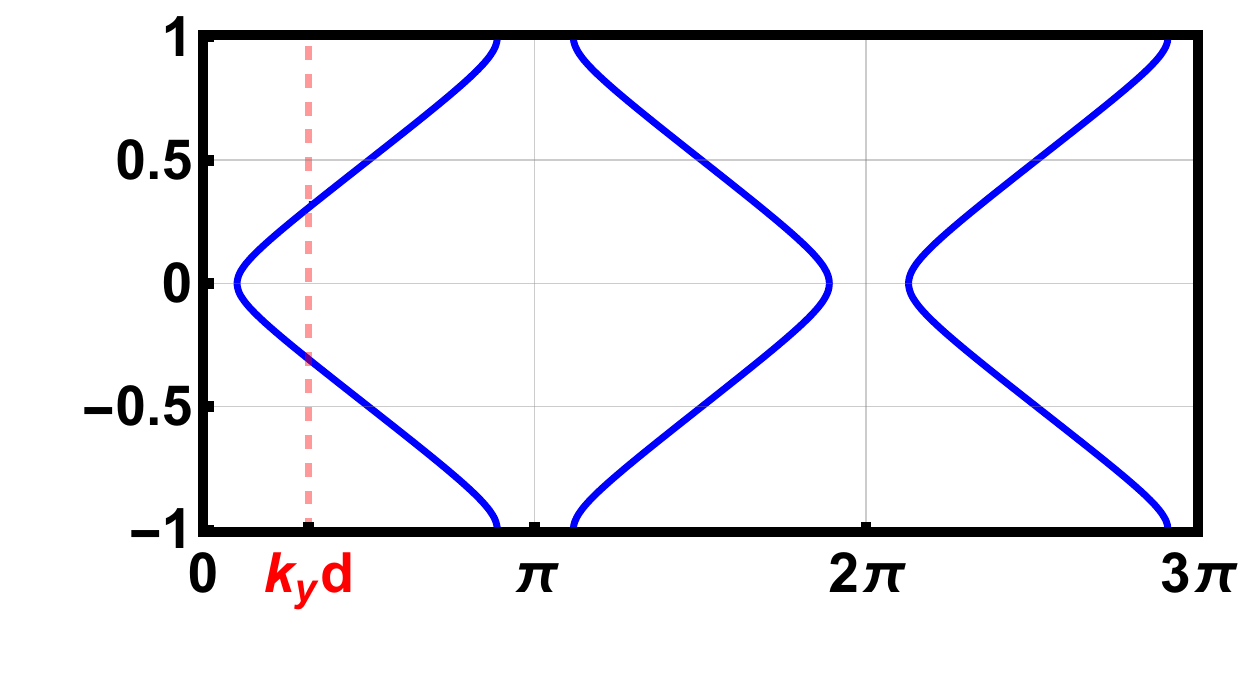}
 		\label{FigKx:SubFigB}
 	}\hspace{-0.51cm}
 	\subfloat[$\mathbb{V}=5\pi$]{
 		\includegraphics[width=5.5cm]{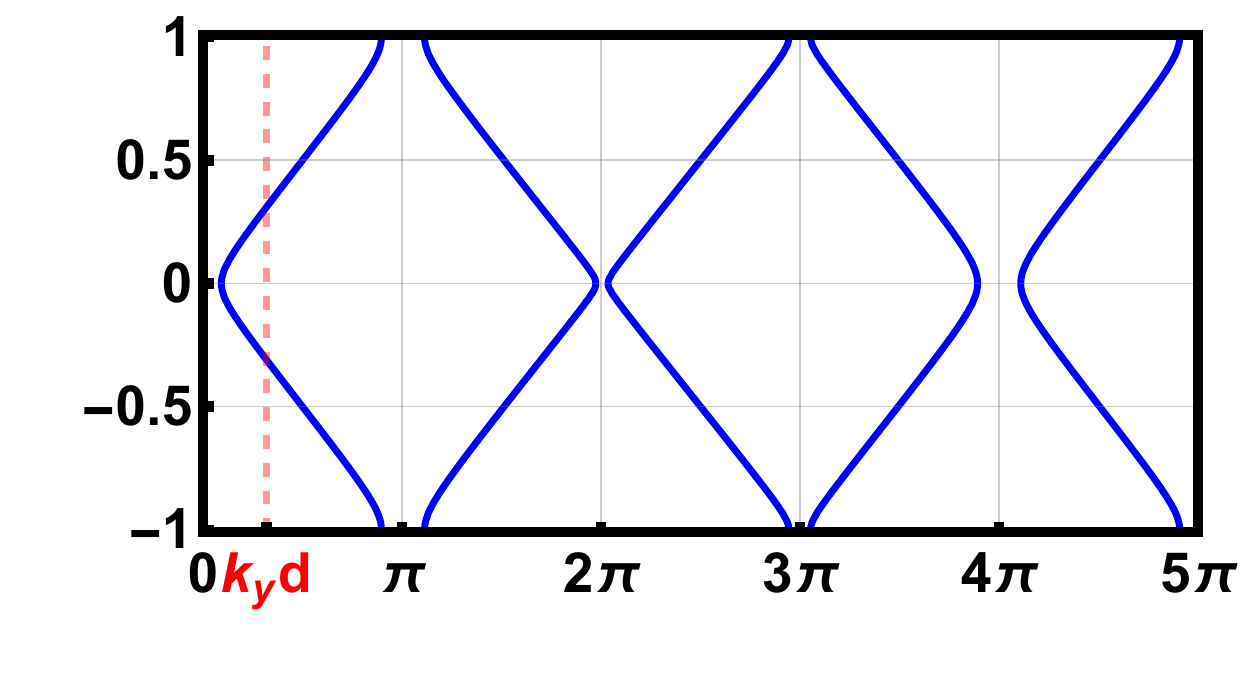}
 		\label{FigKx:SubFigC}
 	}\vspace{-0.7cm}
	\subfloat[$\mathbb{V}=\pi$]{
		\includegraphics[width=5.5cm]{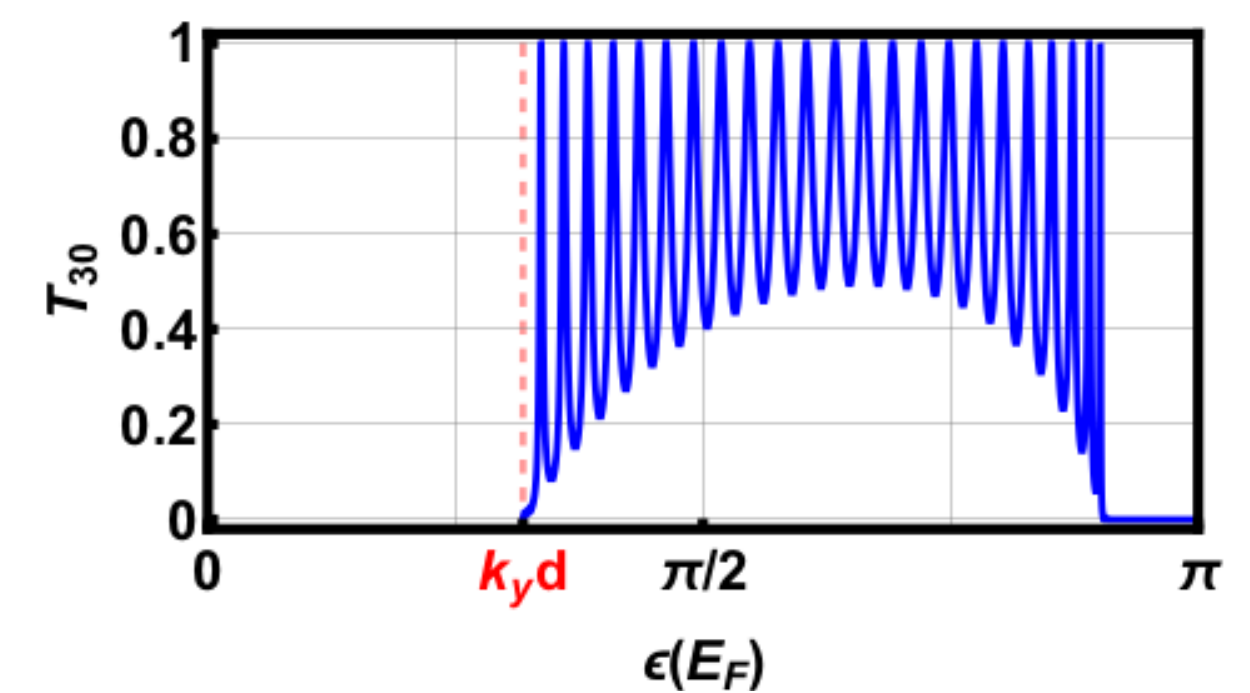}
		\label{FigTKx:SubFigD}
    }\hspace{-0.51cm}
 	\subfloat[$\mathbb{V}=3\pi$]{
 		\includegraphics[width=5.5cm]{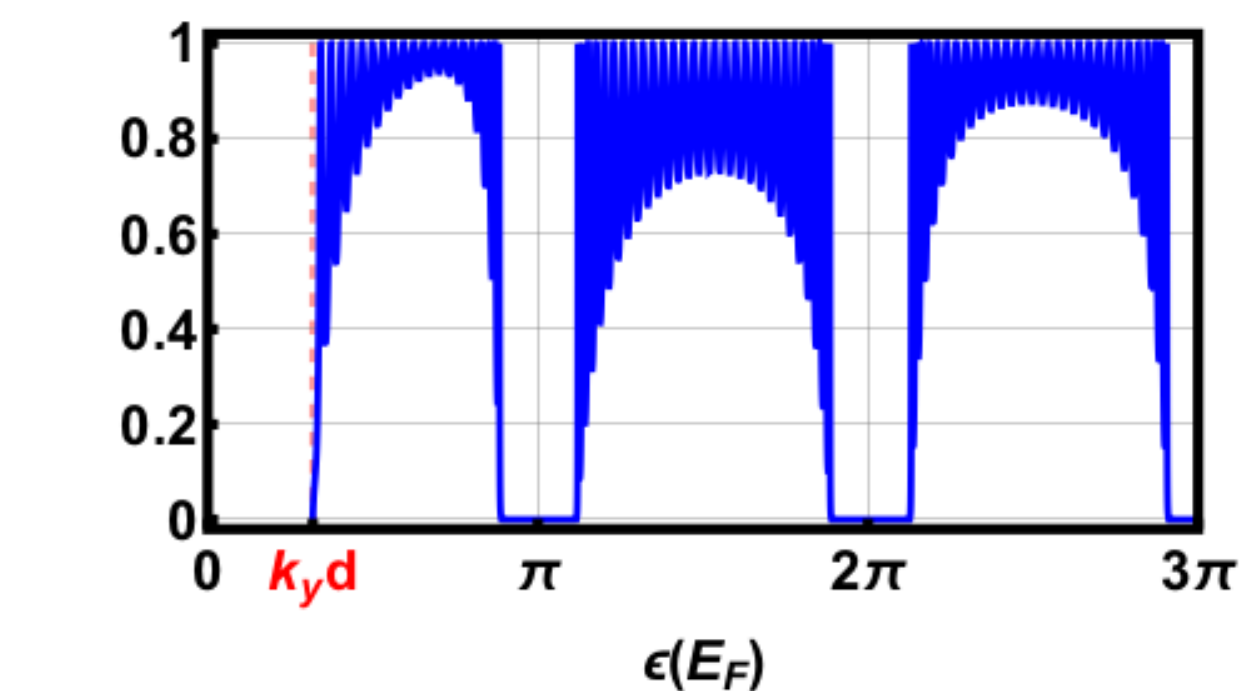}
 		\label{FigTKx:SubFigE}
 	}\hspace{-0.51cm}
 	\subfloat[$\mathbb{V}=5\pi$]{
 		\includegraphics[width=5.5cm]{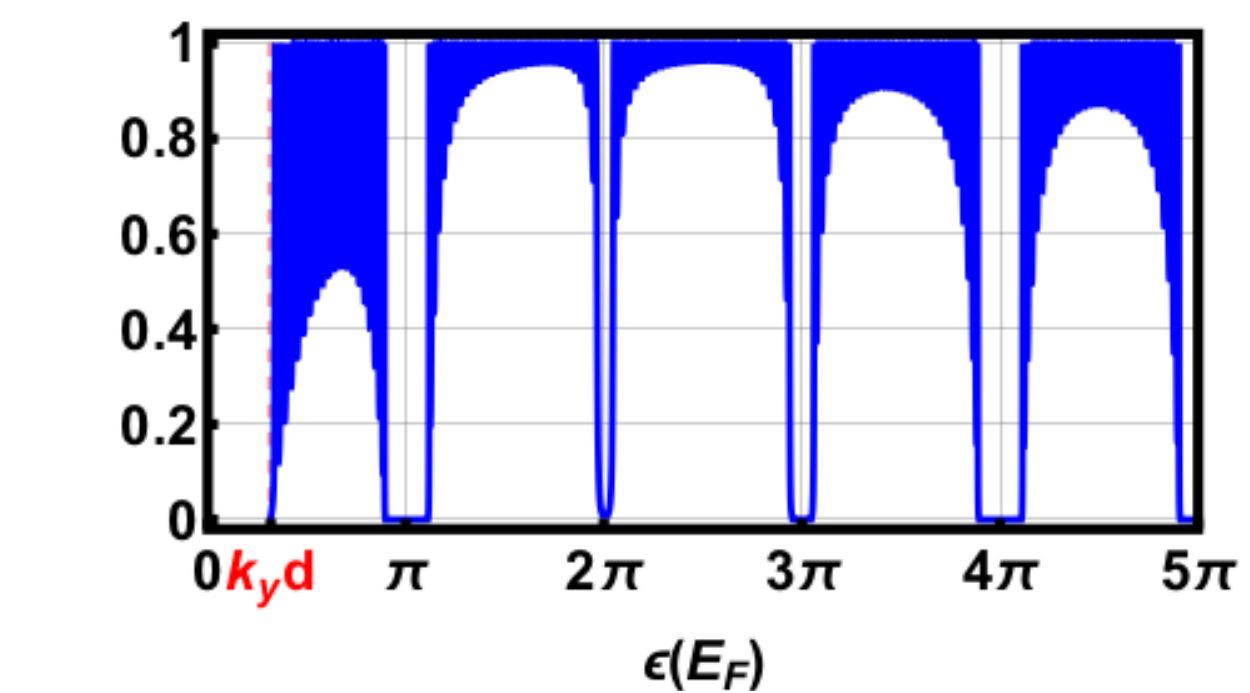}
 		\label{FigTKx:SubFigF}
 	}
 	\caption{
 		(Color online) Electronic band structures for SSLGSL-3R and the
 		corresponding transmissions probability $T_{30}$ versus incident energy $\varepsilon$,
 		with $k_yd=1$, $q_2=\frac{1}{3}$, $n=30$
 		and three values $\mathbb{V}= \pi, 3\pi, 5\pi$ .
 	}
 	\label{FigKx}
 \end{figure}

Figure \ref{FigKx} shows the dispersion relation and the corresponding transmission probability $T_{30}$
versus incident energy $\varepsilon$ for $n=30$, $k_yd=1$
and three values $\mathbb{V}= \pi, 3\pi, 5\pi$.
In
 Figures \ref{FigKx}\subref{FigKx:SubFigA}, \ref{FigKx}\subref{FigKx:SubFigB}, \ref{FigKx}\subref{FigKx:SubFigC}
we observe that each electronic band presents minibands  separated from each other by the band gaps
located at energies $\varepsilon=m\pi$ with different gap widths. Figures
 \ref{FigKx}\subref{FigTKx:SubFigD}, \ref{FigKx}\subref{FigTKx:SubFigE}, \ref{FigKx}\subref{FigTKx:SubFigF} show that
 Dirac fermions have zero transmission for energies coinciding with band gaps in electronic band,
 except for the first near the original Dirac point (ODP) 
 where the transmission is zero up to the value $\varepsilon=k_y d$. For the range $0<\varepsilon<k_y d$, 
 we have bound states  for all barrier height $\mathbb{V}$ and therefore the transmission is null. 
 Now for the second range 
 $k_y d\leqslant \varepsilon\leqslant\mathbb{V}=m\pi$, there is  $m$ transmission gaps,
 which are exactly inside of the $m$ vertical Dirac points (VDPs) 
 enumerated in our previous work \cite{kamal2018}. It is clearly seen that
 the oscillations of $T_{30}$ change  and become condensated
 as long as $\mathbb{V}$ increases.

\begin{figure}[!ht]\centering
    \subfloat[$q_2=0$]{
		\includegraphics[height=4.5cm]{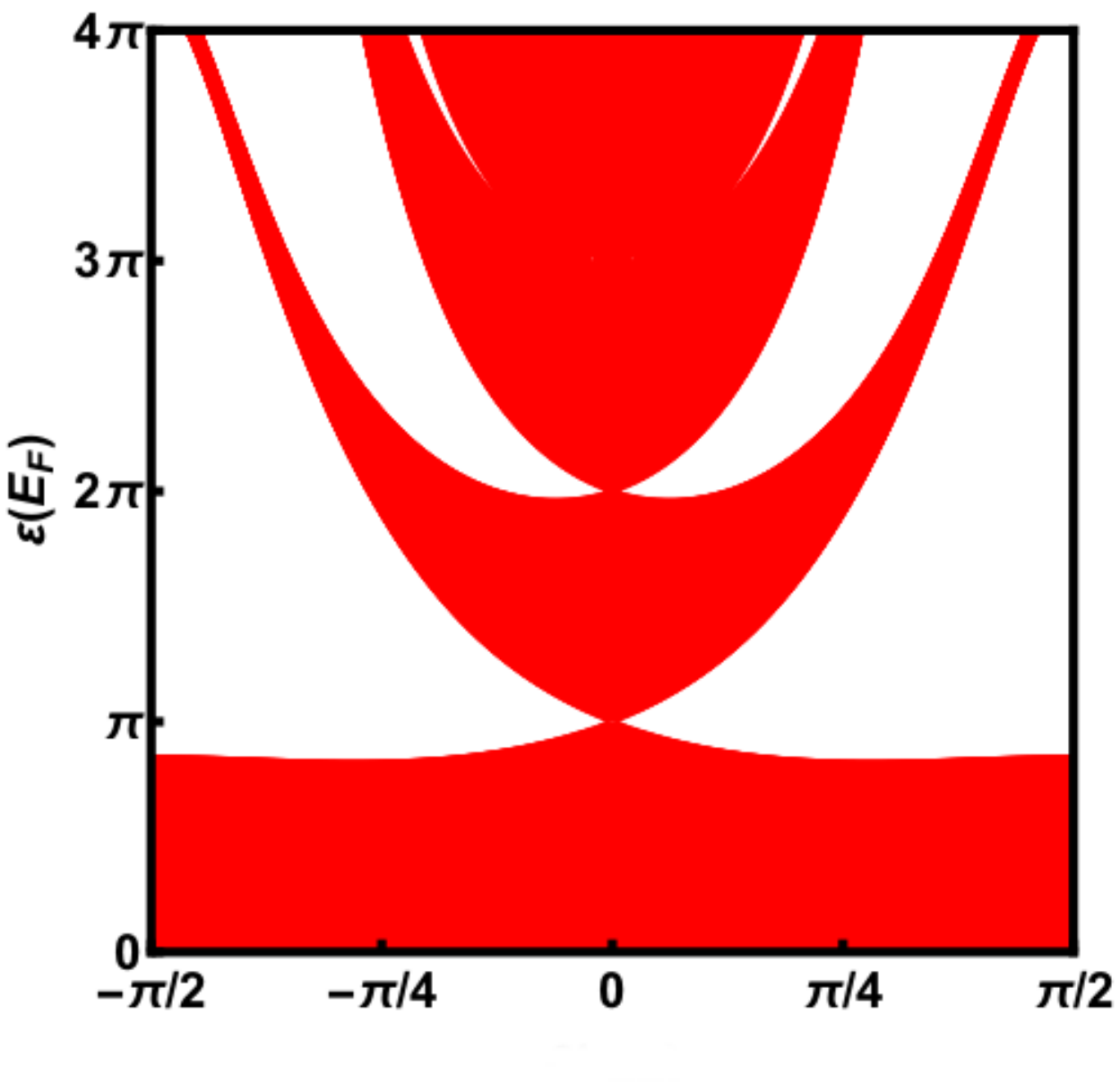}
		\label{FigBS:SubFigC}
	}
	\subfloat[$q_2=0$]{
		\includegraphics[height=4.5cm]{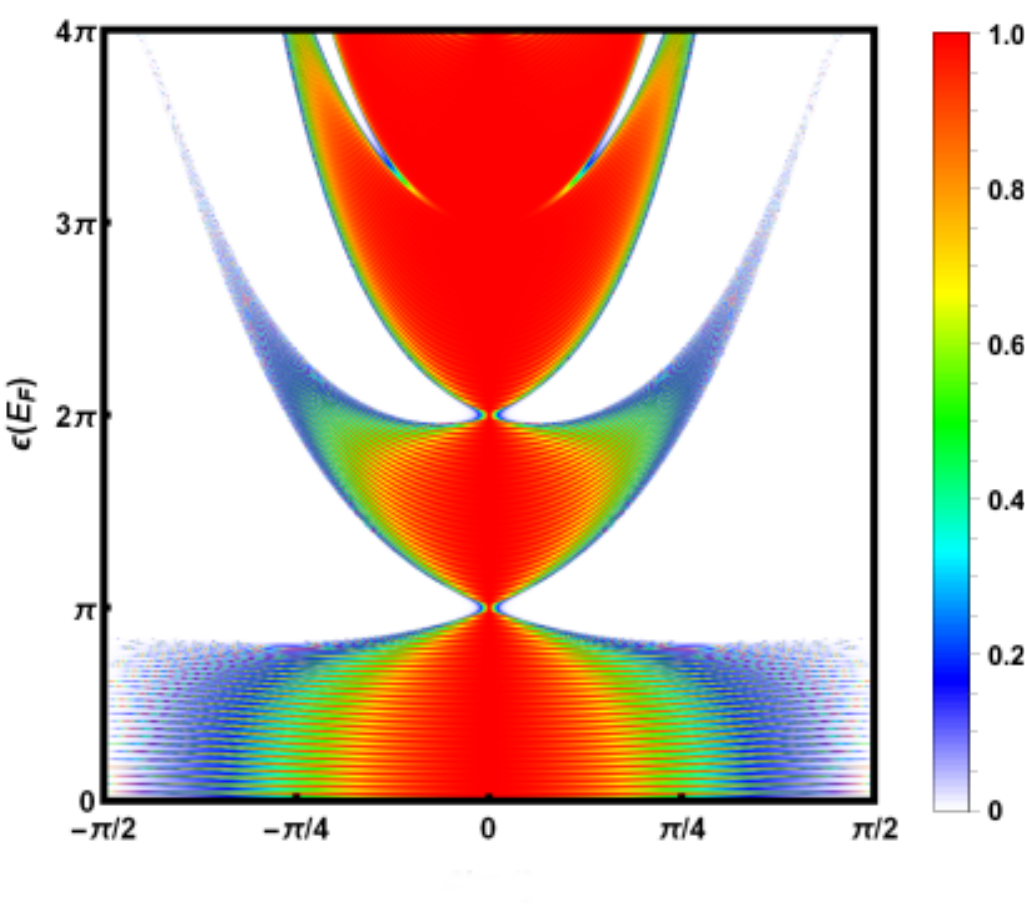}
		\label{FigTtheta:SubFigD}
	}\vspace{-0.5cm}
	\subfloat[$q_2=\frac{1}{3}$]{
		\includegraphics[height=4.5cm]{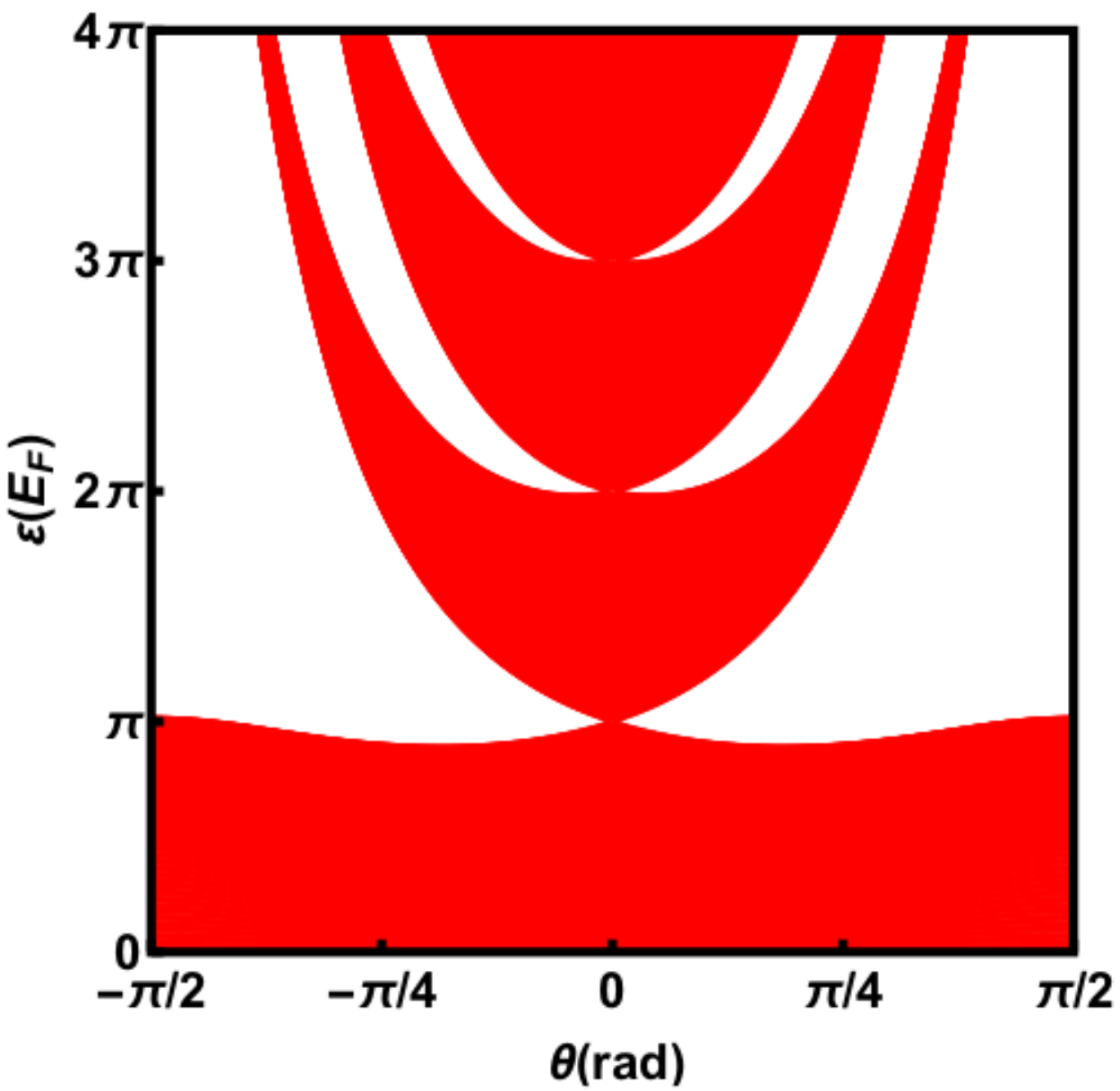}
		\label{FigBS:SubFigA}
	}
	\subfloat[$q_2=\frac{1}{3}$]{
		\includegraphics[height=4.5cm]{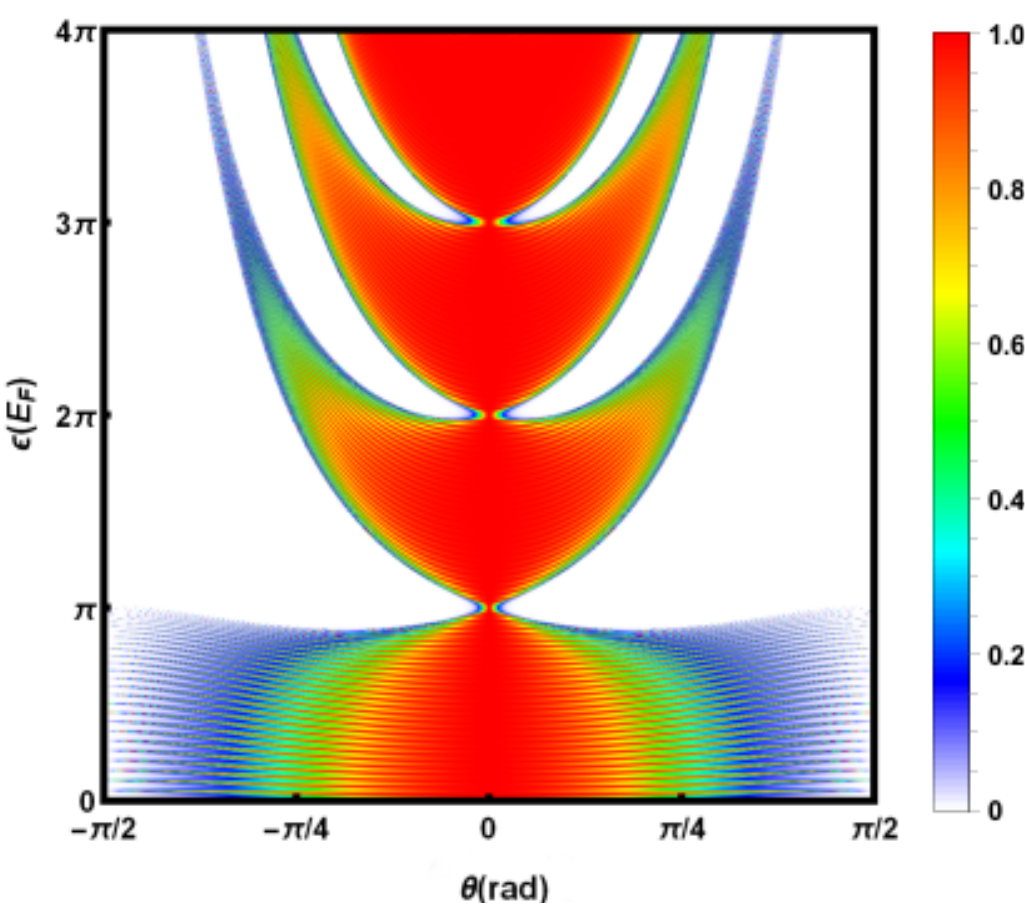}
		\label{FigTtheta:SubFigB}
	}
	\caption{
		(Color online) Density plots of electronic band structures 
		$k_x\in[0,\frac{\pi}{d}]$  and the corresponding transmission probability $T_{30}$ versus
		incident energy $\varepsilon$ and incident angle $\theta$ with $n=30$, $\mathbb{V}=\pi$.	$
 (\protect\subref{FigKx:SubFigA},\protect\subref{FigTtheta:SubFigD}):
 q_2=\frac{1}{3} $,
  $
 (\protect\subref{FigKx:SubFigC},\protect\subref{FigTtheta:SubFigB}):
 q_2=\frac{1}{3}$.
	}
	\label{StructBandTtheta}
\end{figure}

In Figure \ref{StructBandTtheta}, we present density plots of electronic band 
structures and transmission probability $T_{30}$ as
function of incident energy $\varepsilon$ and incident angle $\theta$ where
$k_yd=\varepsilon\sin\theta$, with $n=30$, $\mathbb{V}=\pi$ and for cases
$q_2=0,1/3$. In Figures 
\ref{StructBandTtheta}\subref{FigBS:SubFigC}, \ref{StructBandTtheta}\subref{FigBS:SubFigA}
we observe 
that there are VDPs located
at $\varepsilon=m\pi$ in the dispersion relation as it was point  out in \cite{kamal2018}. In
Figures \ref{StructBandTtheta}\subref{FigTtheta:SubFigD}, \ref{StructBandTtheta}\subref{FigTtheta:SubFigB}
 we see that when $\theta$ increases the width
of each transmission gap increases, where the position of its center is the position of the
corresponding VDPs located at
$\varepsilon=m\pi$ with $m\in\mathbb{Z}$, and the adjacent transmission gaps are separated
by $\Delta\varepsilon=\pi$. When $\theta$ goes to $\pm\frac{\pi}{2}$, with fixed $q_2$ and $\mathbb{V}=\pi$,
the transmission gap is very large. We have a symmetry between positive and negative angles,
and then in the forthcoming analysis we will be interested only on the positive  ones. 
In Figure \ref{StructBandTtheta},
for energies of the band structures lower than that of the first VDP ($\varepsilon\leqslant\pi$),
one has transmission for all angles and consequently the superlattice behaves like a more refractive medium
than that of the pristine graphene. For  energies
$\pi\leqslant\varepsilon\leqslant2\pi$, the reflection is total from a critical angle corresponding to
the boundary between the energy band and the first gap where
all critical angles in this region take
parabolic forms. Beyond the second VDP, other gaps at the level of each VDP with other angles 
that separate energy bands and gaps, we have $(2m-1)$ angles between two consecutive VDPs $m$ and $(m+1)$. 
The change of $q_2$ causes variation of the gap band widths
and there is a decrease in the opening of the parabolas when $q_2$ increases, 
which shows that critical angles can be controlled by tuning on  distance $q_2$.

\begin{figure}[!htb]\centering
	\subfloat[$q_2=0$]{
		\includegraphics[height=4.5cm]{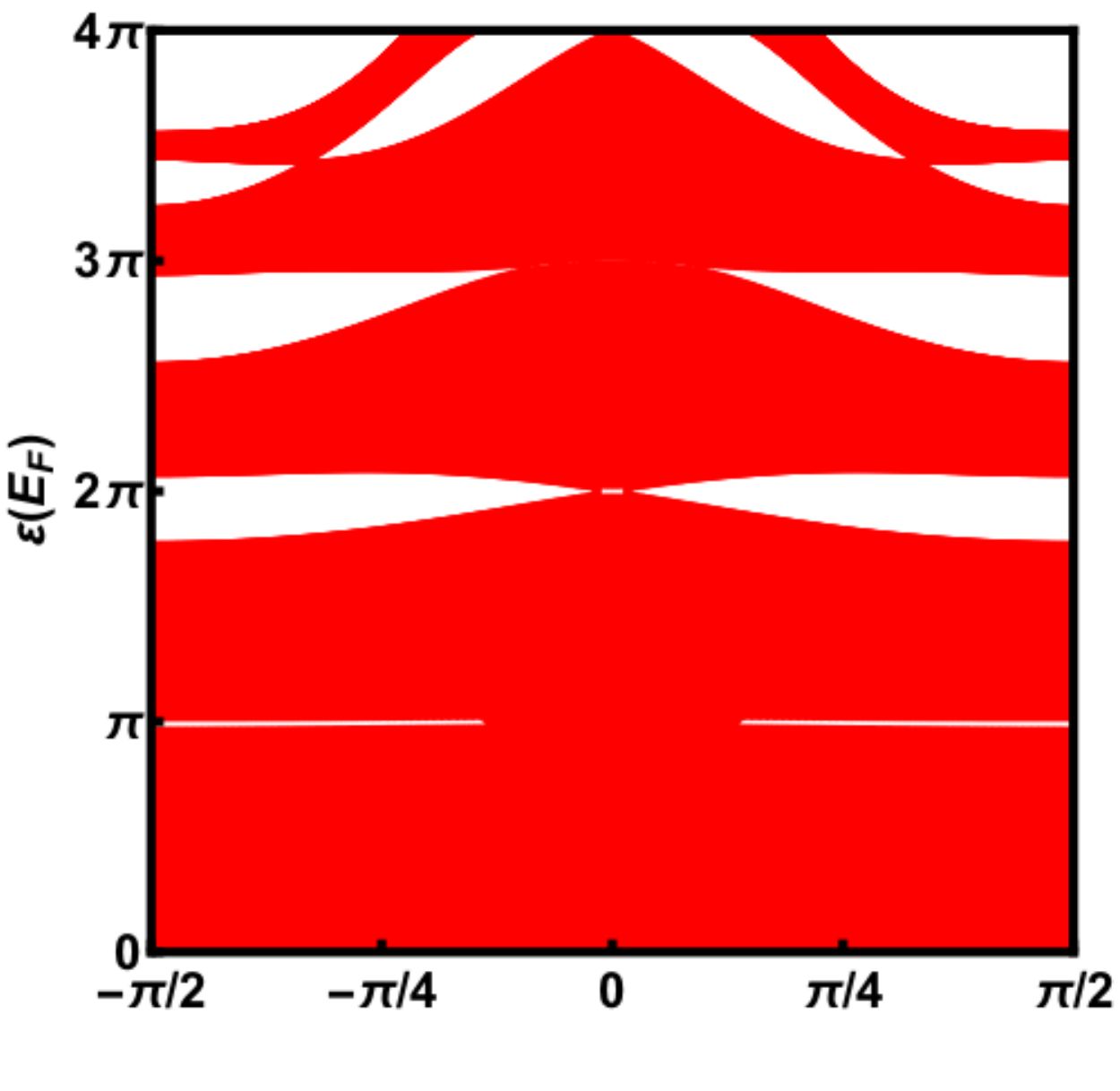}
		\label{FigBSVq0:SubFigA}
	}
	\subfloat[$q_2=0$]{
		\includegraphics[height=4.5cm]{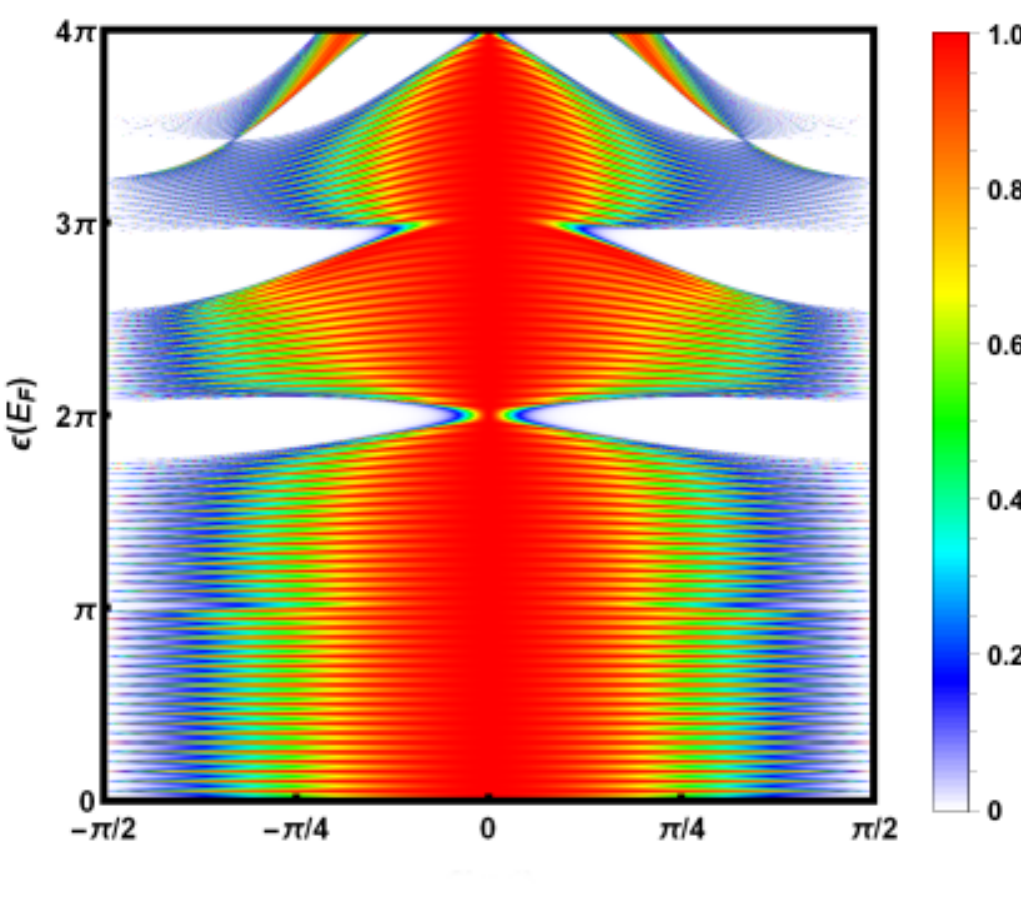}
		\label{FigBSVq0:SubFigB}
	}\vspace{-0.5cm}
		\subfloat[$q_2=\dfrac{1}{3}$]{
		\includegraphics[height=4.5cm]{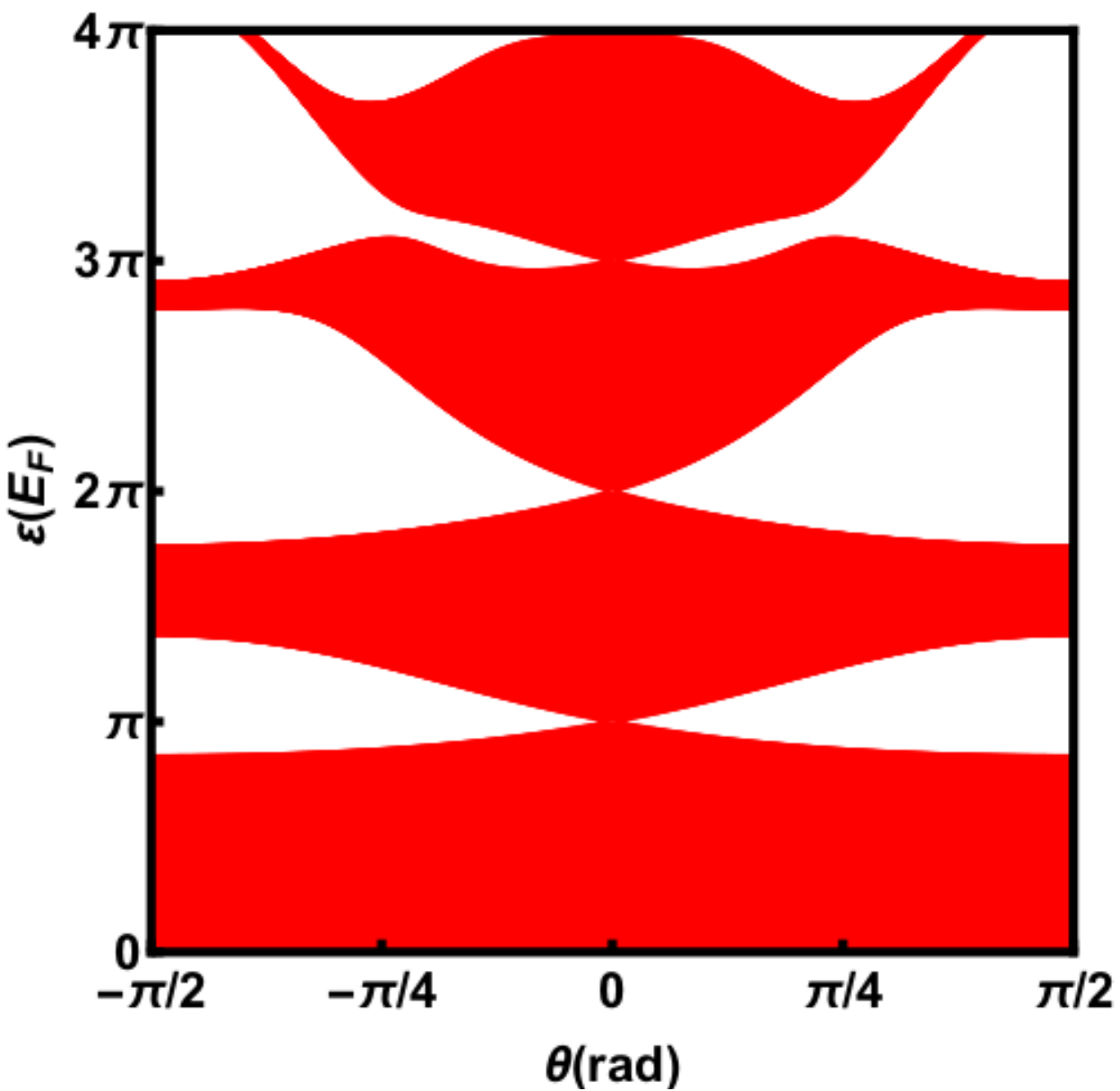}
		\label{FigBSVq13:SubFigC}
	}
	\subfloat[$q_2=\dfrac{1}{3}$]{
		\includegraphics[height=4.5cm]{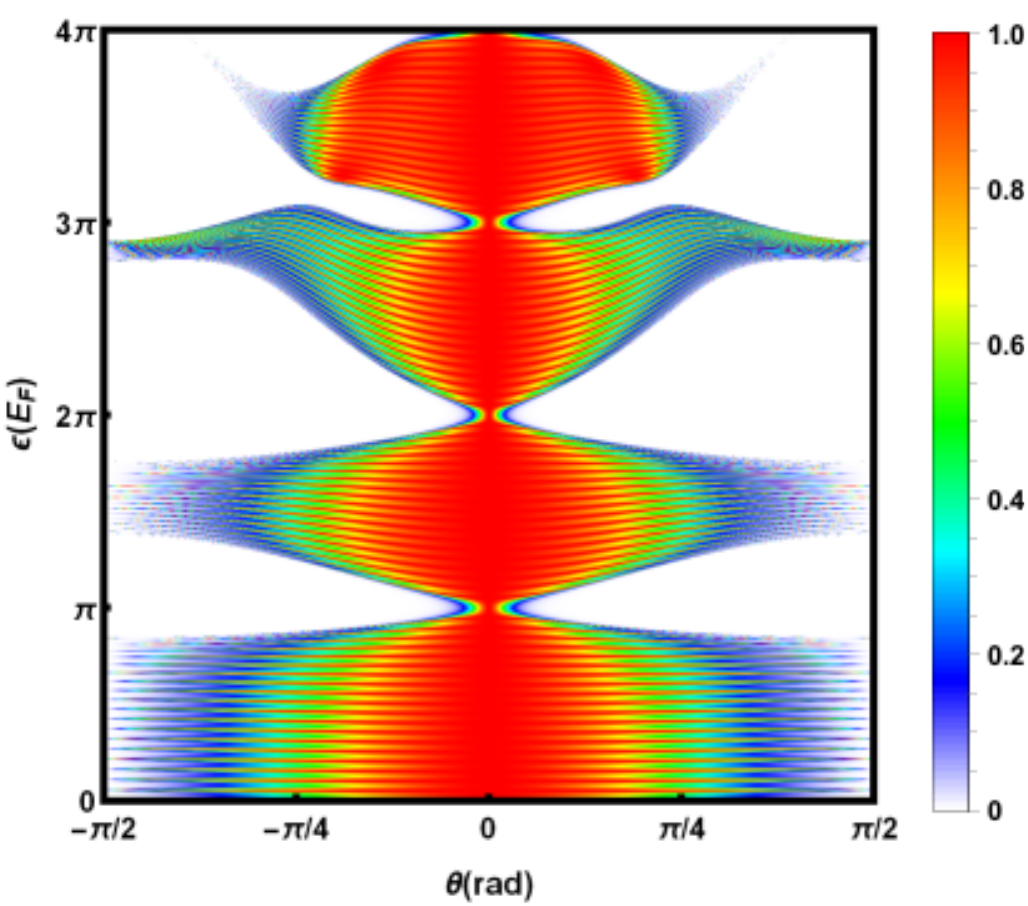}
		\label{FigBSVq13:SubFigD}
	}
	\caption{
		(Color online)
		Density plots of electronic band structure 
		$k_x\in[0,\frac{\pi}{d}]$ 
		(\protect\subref{FigBSVq0:SubFigA}, \protect\subref{FigBSVq13:SubFigC}) 
		and the corresponding transmission probability $T_{30}$ (\protect\subref{FigBSVq0:SubFigB},
		\protect\subref{FigBSVq13:SubFigD})
		versus the incident energy $\varepsilon$ and incident angle 
		$\theta$ with $n=30$, $\mathbb{V}=7\pi$, $q_2=0,1/3$.
	}
	\label{StructBandThetaVq0q13}
\end{figure}

Now let us increase the barrier height to $\mathbb{V}=7\pi$ under the same conditions
as has been done in Figure \ref{StructBandTtheta} with $\mathbb{V}=\pi$. This gives the density plots presented
%
in Figure \ref{StructBandThetaVq0q13}
where we observe that the locations of the VDPs always remains at the same energy values
$\varepsilon=m\pi$, the band gaps start as before near the VDPs but their shapes change compared to
$\mathbb{V}=\pi$ (Figure \ref{StructBandTtheta}). Indeed, 
the parabolic forms of the critical angles disappear by generating energy bands, which
cover all the incident angles
and the widths of allowed zone became large. 
In Figures \ref{StructBandThetaVq0q13}\subref{FigBSVq0:SubFigA}
and \ref{StructBandThetaVq0q13}\subref{FigBSVq0:SubFigB}, the band gaps are narrow  compared to those
observed in Figures \ref{StructBandThetaVq0q13}\subref{FigBSVq13:SubFigC} and
\ref{StructBandThetaVq0q13}\subref{FigBSVq13:SubFigD}, also we notice that their number decreases. In the range 
$3\pi \leq \varepsilon \leq 4\pi$ and $q_2=0$, 
a new Dirac point is generated and located between
two incident angles $\theta=\dfrac{\pi}{4}$ and $\theta=\dfrac{\pi}{2}$.
However such  point is eliminated for the case $q_2=\dfrac{1}{3}$  and   an opening gap appeared 
at its place. We see that changing the barrier height the behavior of electronic band structures
and transmission changes completely from
Figure \ref{StructBandTtheta} to Figure \ref{StructBandThetaVq0q13}. This change tells us that
we can control both transmission probability by varying such height.




\begin{figure}[!htb]\centering
    \subfloat[$\varepsilon=\pi$]{
        \includegraphics[height=4.5cm]{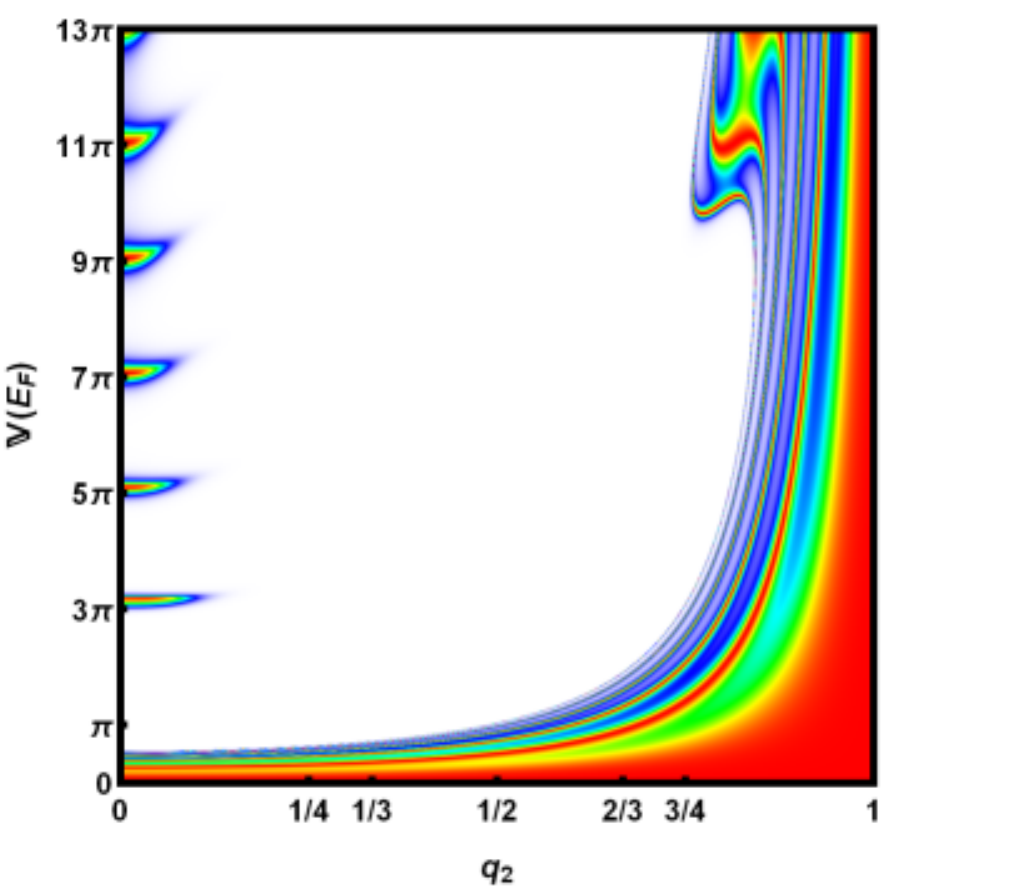}
        \label{FigTq2:SubFigA}
    }\hspace{-0.8cm}
    \subfloat[$\varepsilon=2\pi$]{
        \includegraphics[height=4.5cm]{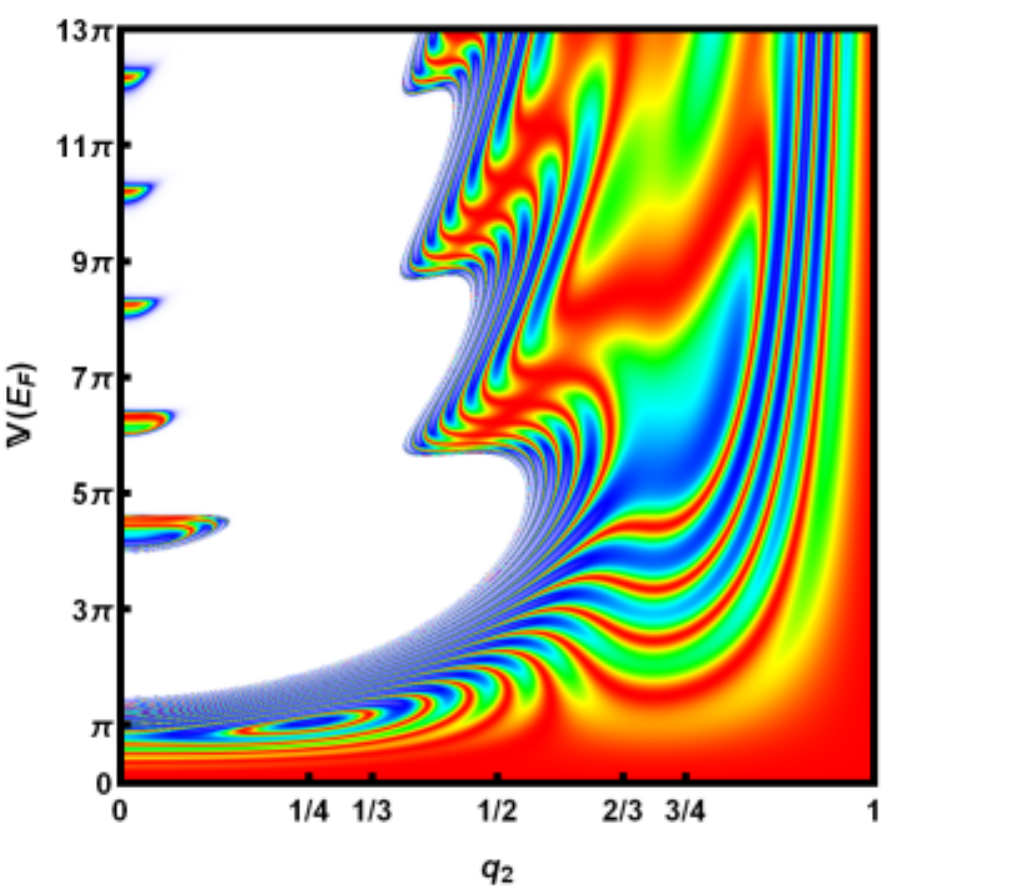}
        \label{FigTq2:SubFigB}
    }\hspace{-0.8cm}
	\subfloat[$\varepsilon=3\pi$]{
        \includegraphics[height=4.5cm]{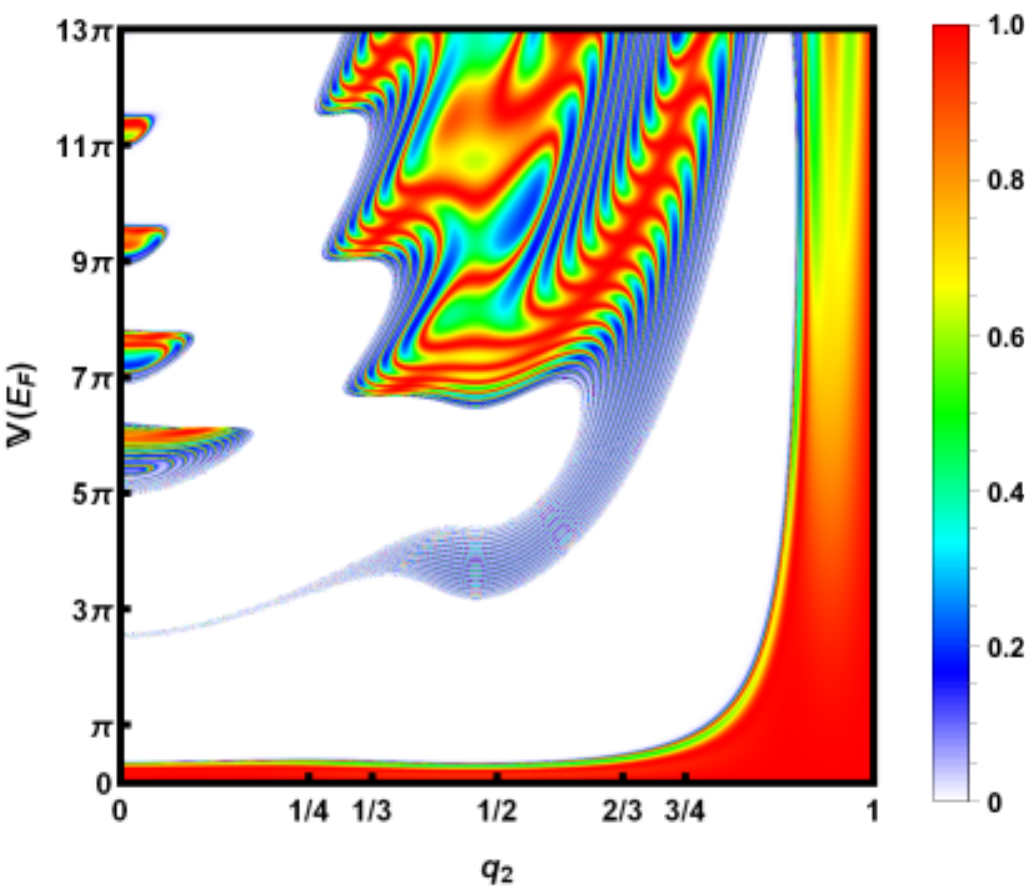}
        \label{FigTq2:SubFigC}
    }
    \caption{
		(Color online) Density plot of transmission probability $T_{30}$ versus
		distance $q_2$ and potential $\mathbb{V}$, 
		with $n=30$, $\theta=\frac{\pi}{4}$, $\varepsilon=\pi,~2\pi,~3\pi$.
	}
	\label{FigTq2}
\end{figure}

Figure \ref{FigTq2} shows density plots of transmission probability $T_{30}$ as function
of potential $\mathbb{V}$ and distance $q_2$ with $n=30$, $\theta=\dfrac{\pi}{4}$, 
for three values of incident energy $\varepsilon=\pi,~2\pi,~3\pi$. 
In the case of pristine graphene which can be realized either by $q_2=1$ or
$\mathbb{V}=0$, we have a total transmission (red color). 
In the case of SSLGSL-2R ($q_2=0$), we have transmission bands intercalated alternately by gaps.
By moving away from pristine graphene ($\mathbb{V}=0$) increasing either  $\mathbb{V}$,
the energy bands are gradually eliminated and appear in the form of lobes (clusters) surrounded by band gaps,
the dimensions of these lobes depend on both $\varepsilon$ and $\mathbb{V}$. 
The band gap is spread over 
until a band of transmission located on the right hand in  Figure \ref{FigTq2}. This last
transmission band has resonance peaks and bumps in its left one for $\varepsilon=\pi$, three for
$\varepsilon=2\pi$, four for $\varepsilon=3\pi$. It is interesting to note that
in Figure \ref{FigTq2}\subref{FigTq2:SubFigC},
the transmission band split into two bands separated by a gap band and when the energy increases
the lobes move upwards.

\begin{figure}[!ht]\centering
    \subfloat[$\mathbb{V}=3\pi$]{
        \includegraphics[height=4.5cm]{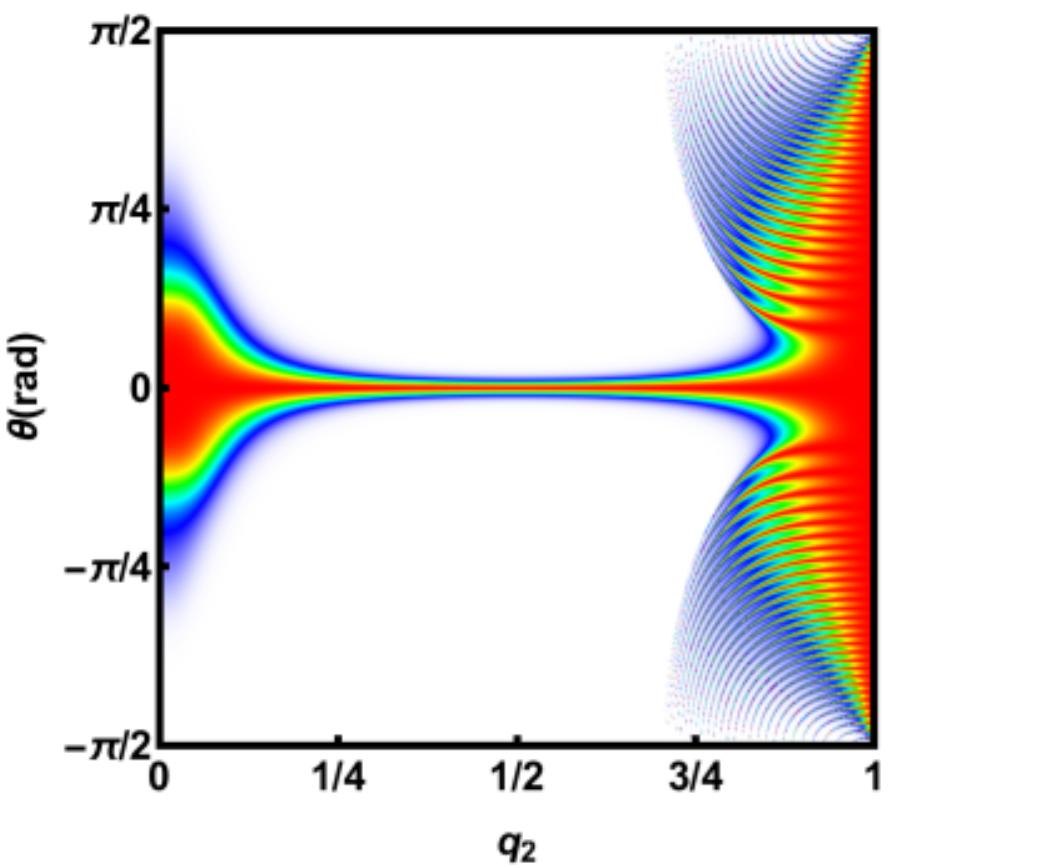}
        \label{FigTrq2Theta:SubFigA}
    }\hspace{-0.8cm}
    \subfloat[$\mathbb{V}=5\pi$]{
        \includegraphics[height=4.5cm]{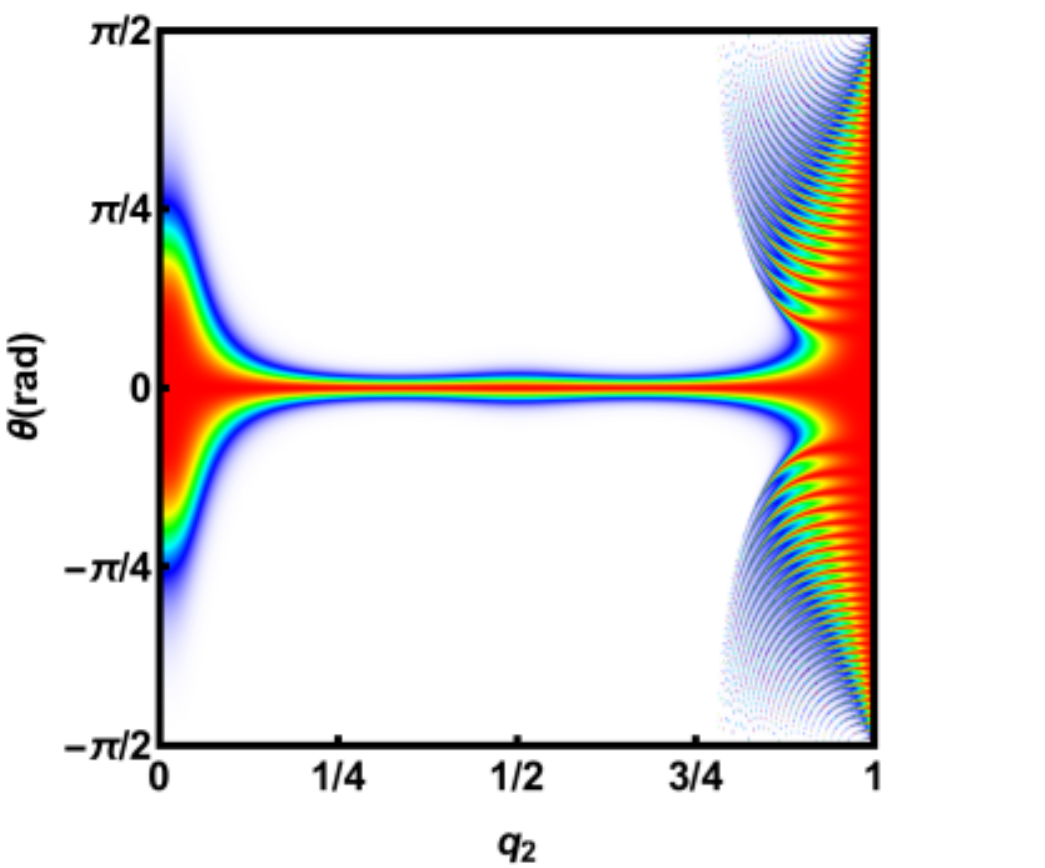}
        \label{FigTrq2Theta:SubFigB}
    }\hspace{-0.8cm}
    \subfloat[$\mathbb{V}=7\pi$]{
        \includegraphics[height=4.5cm]{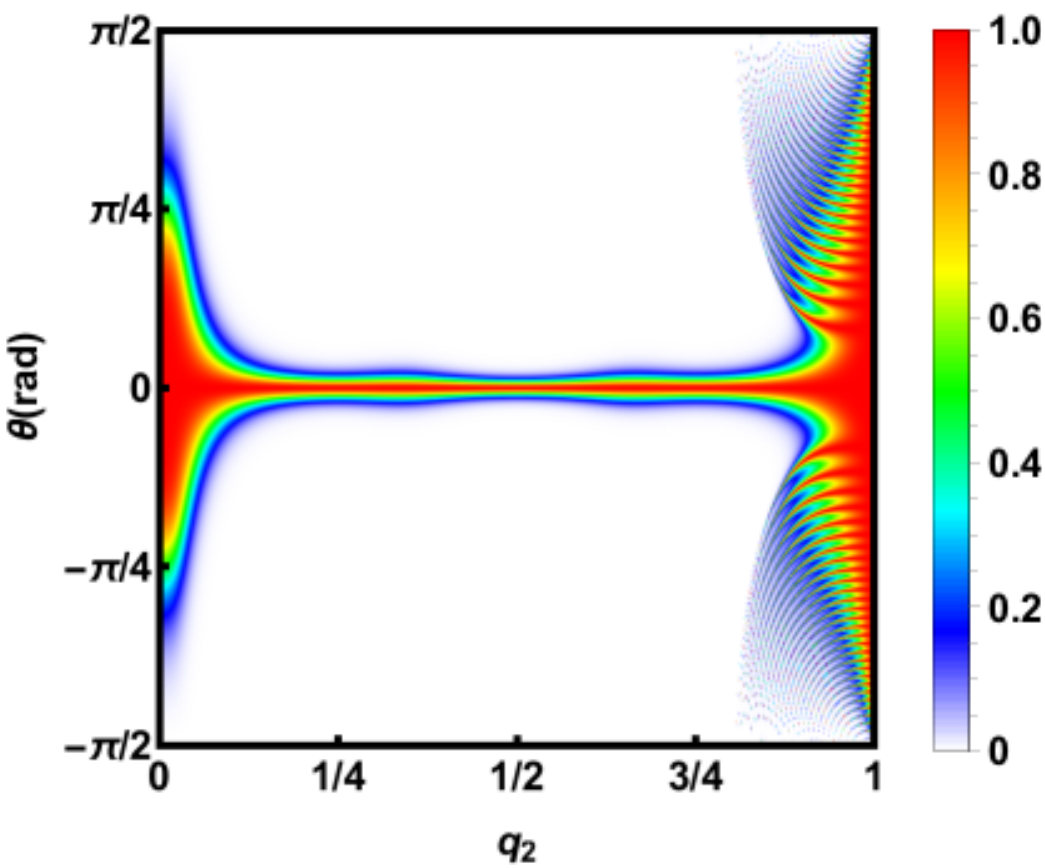}
        \label{FigTrq2Theta:SubFigC}
    }
    \caption{
		(Color online) Density plot of transmission probability $T_{30}$  versus
		incident angle $\theta$ and distance $q_2$, with $n=30$, $\varepsilon=\pi$,
		for \protect \subref{FigTrq2Theta:SubFigA}: $\mathbb{V}=3\pi$,
		\protect \subref{FigTrq2Theta:SubFigB}: $\mathbb{V}=5\pi$, \protect \subref{FigTrq2Theta:SubFigC}:
		$\mathbb{V}=7\pi$.
	}
	\label{FigTrq2ThetaE1pi}
\end{figure}

\begin{figure}[!htb]\centering
    \subfloat[$\varepsilon=2\pi$]{
        \includegraphics[height=4.5cm]{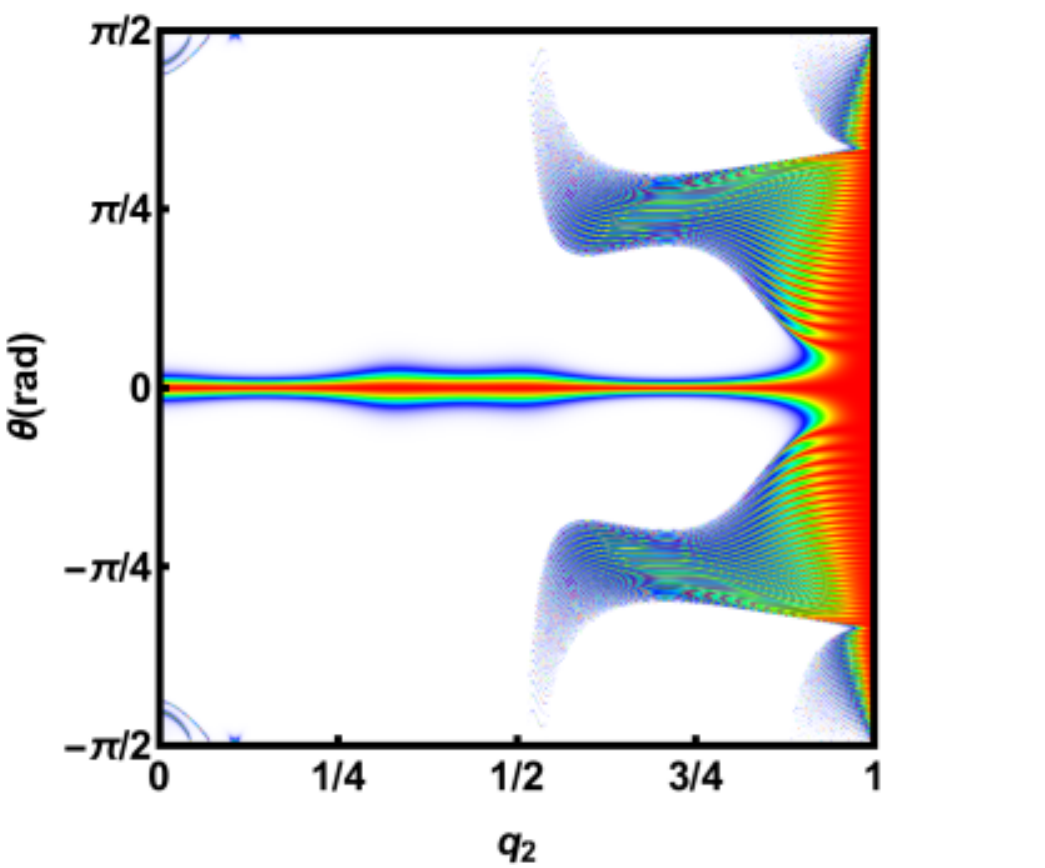}
        \label{FigTrTheta:SubFigB}
    }\hspace{-0.8cm}
    \subfloat[$\varepsilon=3\pi$]{
        \includegraphics[height=4.5cm]{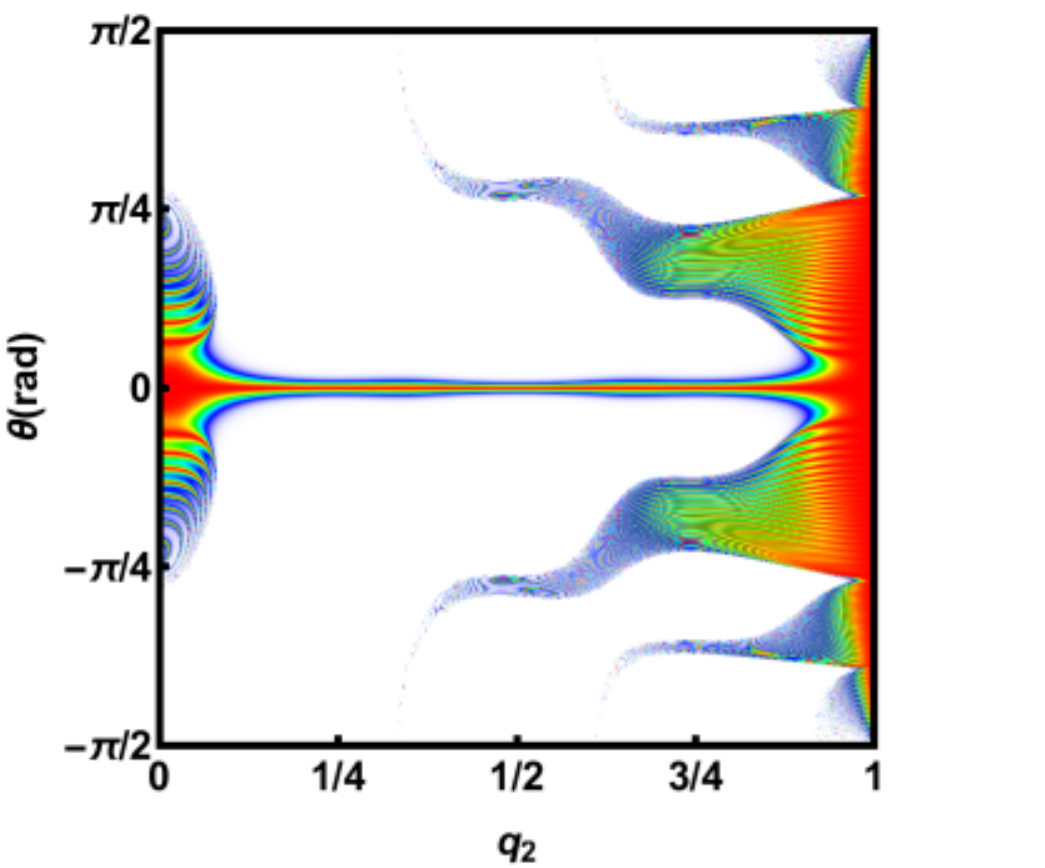}
        \label{FigTrTheta:SubFigC}
    }\hspace{-0.8cm}
    \subfloat[$\varepsilon=4\pi$]{
        \includegraphics[height=4.5cm]{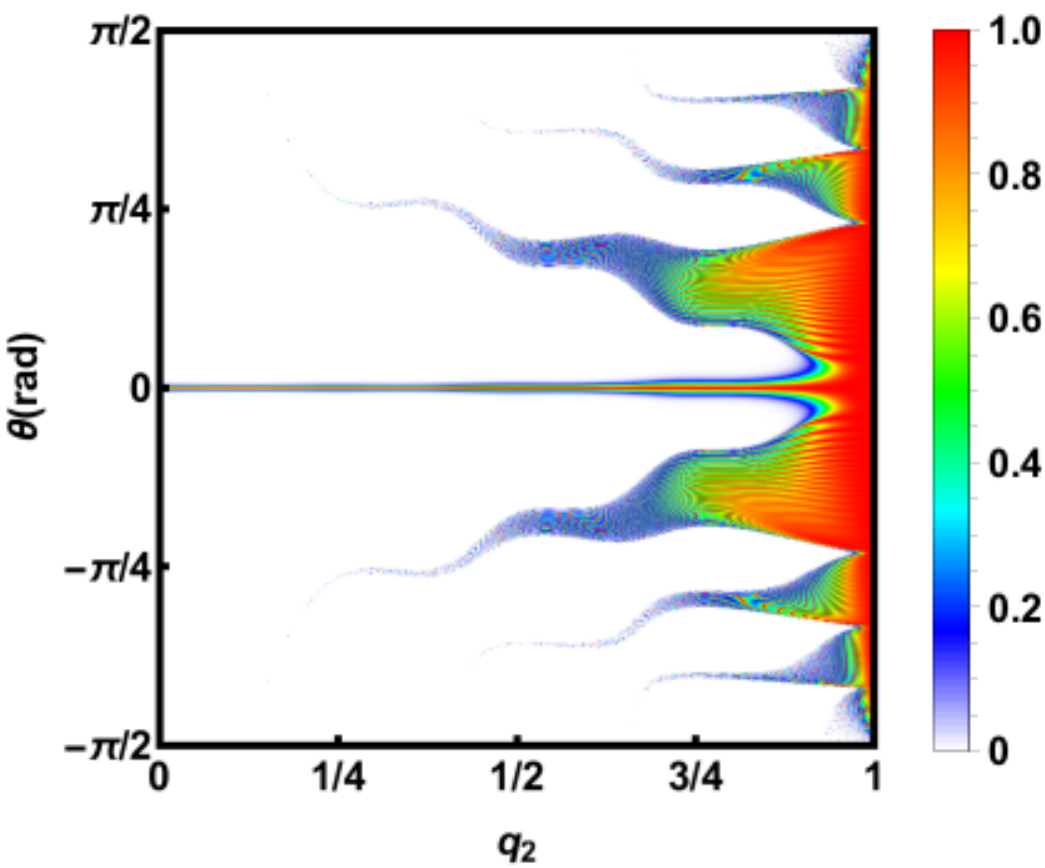}
        \label{FigTrTheta:SubFigD}
    }
    \caption{
		(Color online) Density plot of transmission probability $T_{30}$  versus
		incident angle $\theta$ and distance $q_2$ with $n=30$, $\mathbb{V}=5\pi$,
		for \protect \subref{FigTrTheta:SubFigB}:
		$\varepsilon=2\pi$, \protect \subref{FigTrTheta:SubFigC}: $\varepsilon=3\pi$,
		\protect \subref{FigTrTheta:SubFigD}: $\varepsilon=4\pi$.
	}
	\label{FigTrTheta}
\end{figure}

In Figure \ref{FigTrq2ThetaE1pi}, we present the density plot
of transmission probability as function
of  incident angle $\theta$ and distance $q_2$ 
with $n=30$, $\varepsilon=\pi$ and for three values of potential
$\mathbb{V}=3~\pi,~5\pi,~7\pi$. 
We observe around the normal incident angle $\theta=0$ for all $q_2$,
there is always a total transmission and when $\theta$ increases a bad gap appears
(white color). When $q_2$ is near zero, the transmission band takes place at $\theta=0$
and becomes large as long as the barrier height $\mathbb{V}$ increases. Note that,
existence of transmission gaps around $q_2$  for three values of $\mathbb{V}$ is
justified by bands intercalated by bad gaps at $q_2=2$, see Figure \ref{FigTq2}\subref{FigTq2:SubFigA}. 
Now by increasing $q_2$, one sees  that the width of such band along $\theta$-direction decreases
and becomes constant as well as there is apparition of the boosts one for $\mathbb{V}=5\pi$
and two for $\mathbb{V}=7\pi$. For $q_2\geq 3/4$, the transmission band takes a form
covering all incident angles with resonance peaks.


Figure \ref{FigTrTheta} shows the density plot of
 transmission probability  as function of the incident
angle $\theta$ and distance $q_2$ with $n=30$, $\mathbb{V}=5\pi$ and for three values of
incident energy
$\varepsilon=2\pi,~3\pi,~4\pi$. 
We observe that around the normal incident angle $\theta=0$ for all $q_2$ there is always
total transmission. 
It is clearly seen that
for $\varepsilon=m\pi$ there are $2m$ lateral
bands and when $\varepsilon$ increases these bands tend towards $q_2=0$. For $m$ odd, 
when $q_2$ goes to zero, there is a total transmission even for non-null incident angle exhibiting Klein paradox.

\begin{figure}[!ht]\centering
    \subfloat[]{
        \includegraphics[height=4.5cm]{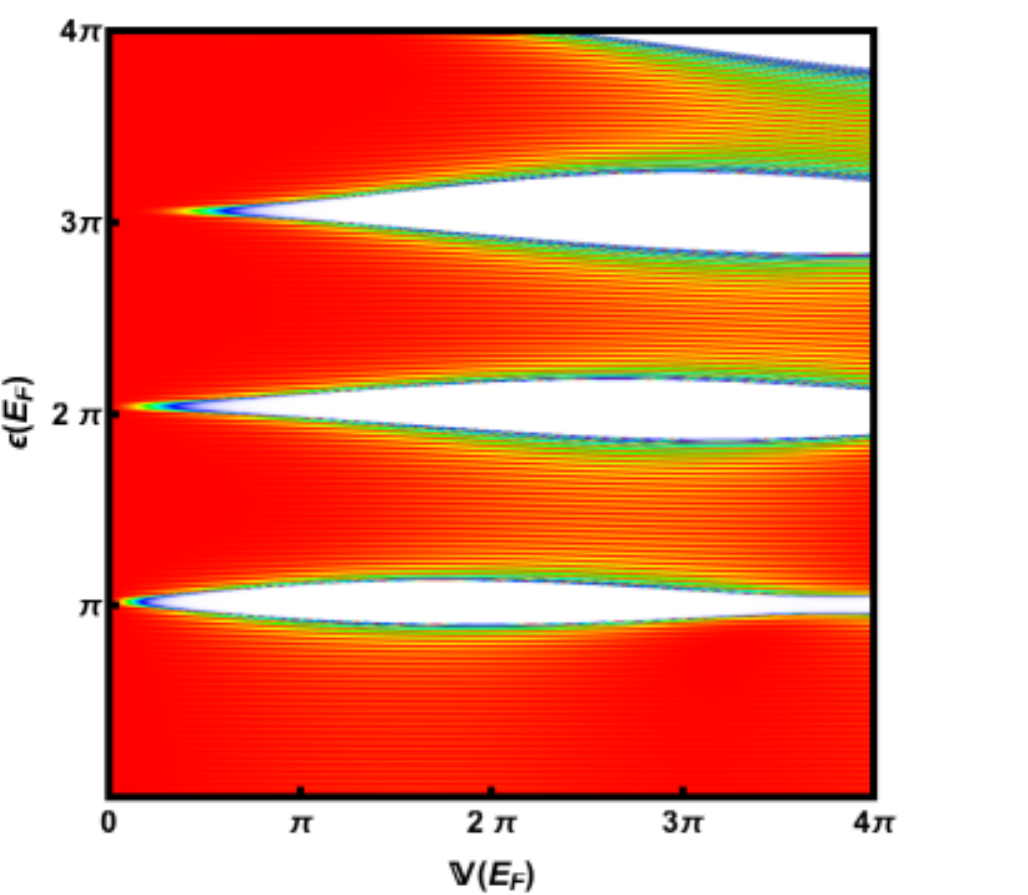}
        \label{FigTthetaVEq:SubFigA}
    }\hspace{-0.8cm}
    \subfloat[]{
        \includegraphics[height=4.5cm]{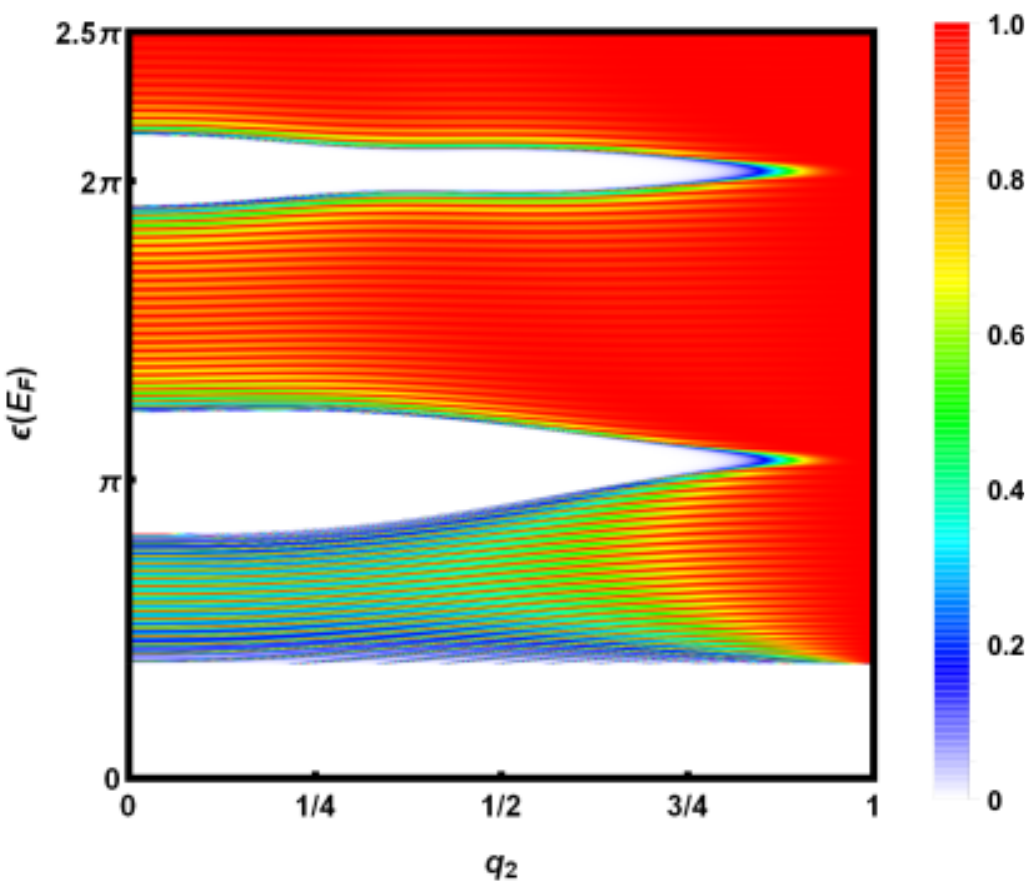}
        \label{FigTthetaVEq:SubFigB}
    }
    \caption{
		(Color online) Density plot of transmission probability $T_{30}$  
		\protect \subref{FigTthetaVEq:SubFigA}: versus incident energy $\varepsilon$
		and potential $\mathbb{V}$, with $q_2=\frac{1}{3}$, $n=30$, $\theta=\frac{\pi}{16}$,
		and \protect \subref{FigTthetaVEq:SubFigB}: versus incident energy $\varepsilon$ and
		distance $q_2$, with $\mathbb{V}=4\pi$, $n=30$, $k_yd=1.2$.
	}
	\label{FigTr:FigTVq2}
\end{figure}

In order to study the effect of the variation of potential $\mathbb{V}$ on the transmission gaps,
we show in Figure \ref{FigTr:FigTVq2}\subref{FigTthetaVEq:SubFigA} the density plot of transmission
probability as function of incident energy $\varepsilon$ and potential $\mathbb{V}$.
We observe that there are energy regions between the VDPs where the incident electrons
can penetrate through the SSLGSL-3R easily, the electrons behave like particles in the free space
almost unfettered. The width of transmission gaps located in VDPs increases as long as $\mathbb{V}$ increases.
We see that when $\mathbb{V}$ is small, the transmission probability is one near VDPs and decreases gradually
until transmission gap appeared.
In Figure \ref{FigTr:FigTVq2}\subref{FigTthetaVEq:SubFigB} we present the density plot of transmission
probability $T_{30}$ for SGSL-3R versus  incident energy $\varepsilon$ and distance $q_2$ with $n=30$.
We choose $\mathbb{V}\!=\!4\pi$ and $k_yd=1.2$ to stay widely near ODP. When $q_2$ varies
from $0$ to $1$, the system gradually changes from a SSLGSL-2R to a pristine graphene through  SSLGSL-3R.
The white regions located at $\varepsilon=\pi,~2\pi$, are transmission gaps near VDPs. The white region
located near the ODP where $\varepsilon=0$, is the transmission gap near this point and has the width of
$|2 k_yd|$, i.e. when the system becomes pristine graphene. When $q_2$ increases, the width of transmission
gaps decreases until its disappearance when the system becomes pristine graphene. When
$q_2\rightarrow 1$, the band structures contain only the ODP
with a purely linear dispersion \cite{kamal2018}. This is in perfect match with result
presented in our previous work
 for band structures of SSLGSL-3R \cite{kamal2018}.

\section{Conductance and Fano factor}

With the transmission probability $T_n$, we can obtain the total conductance $G$
of the system at
zero temperature according to the Landauer B\"uttiker formula \cite{Buttiker1985}. According
to our results, the corresponding 
ballistic conductance under zero temperature is given by
\begin{eqnarray}%
	G&=&(4e^2/h)\int_{-E}^{E}T_n(E,k_y)\frac{dk_y}{2\pi/L_y}\\
	&=&G_0\int_{-\pi/2}^{\pi/2}T_n(\varepsilon,\theta)\cos\theta d\theta\label{Conductance}
\end{eqnarray}
where $L_y$ is the sample size along the $y$-direction, $\theta$ is the incident angle relative
to the $x-$direction, $\theta=\arccos(k_{in} d/\varepsilon)$ and $G_0=2e^2\varepsilon L_y/(\pi h d)$
is the conductance unit. Also we consider 
the Fano factor \cite{Trauzettel2006}, which can be written 
as
\begin{equation}\label{Fano}
	F=\frac{\int_{-\pi/2}^{\pi/2}T_n(1-T_n)\cos\theta d\theta}{\int_{-\pi/2}^{\pi/2}T_n\cos\theta d\theta}.
\end{equation}
These results will be investigated numerically to underline our system behavior. In particular,
we establish  the relationship between two above quantities and the vertical Dirac points (VDPs).

\begin{figure}[!ht]\centering
    \subfloat[$n=1$]{
        \includegraphics[width=7cm]{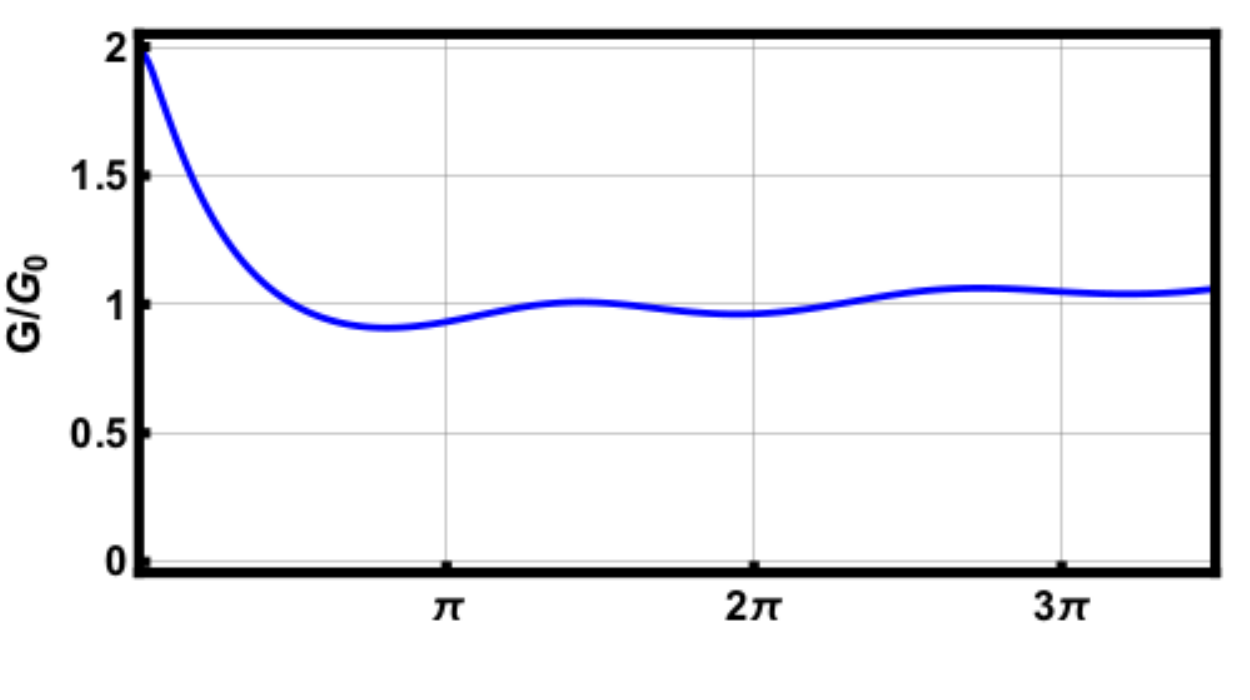}
        \label{FigG:SubFigA}
    }\hspace{-0.3cm}
    \subfloat[$n=10$]{
        \includegraphics[width=7cm]{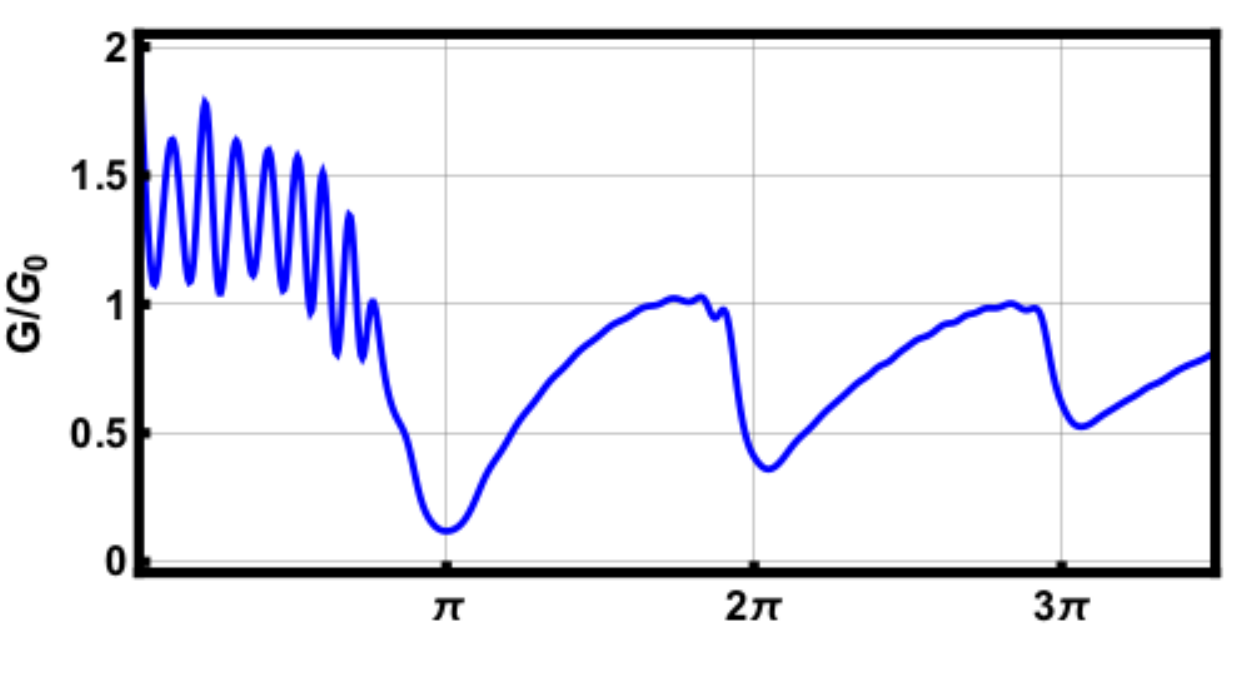}
        \label{FigG:SubFigB}
    }\vspace{-0.5cm}
    \subfloat[$n=20$]{
        \includegraphics[width=7cm]{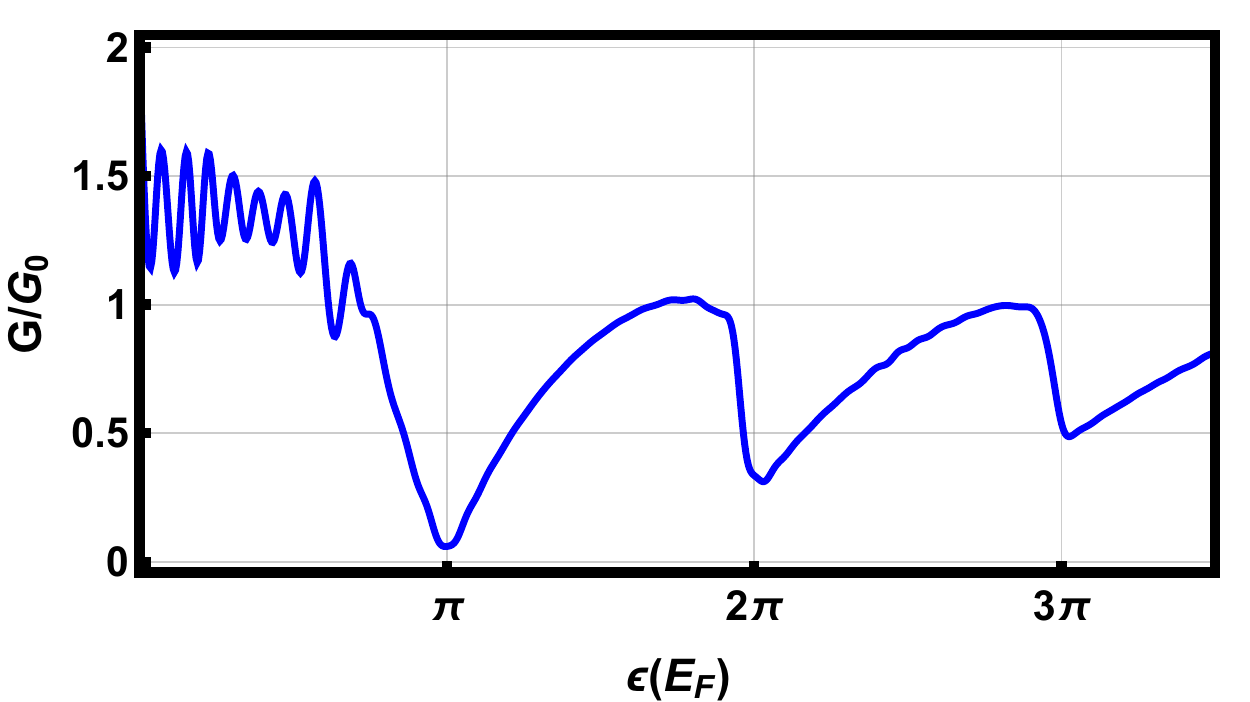}
        \label{FigG:SubFigC}
    }\hspace{-0.3cm}
    \subfloat[$n=30$]{
        \includegraphics[width=7cm]{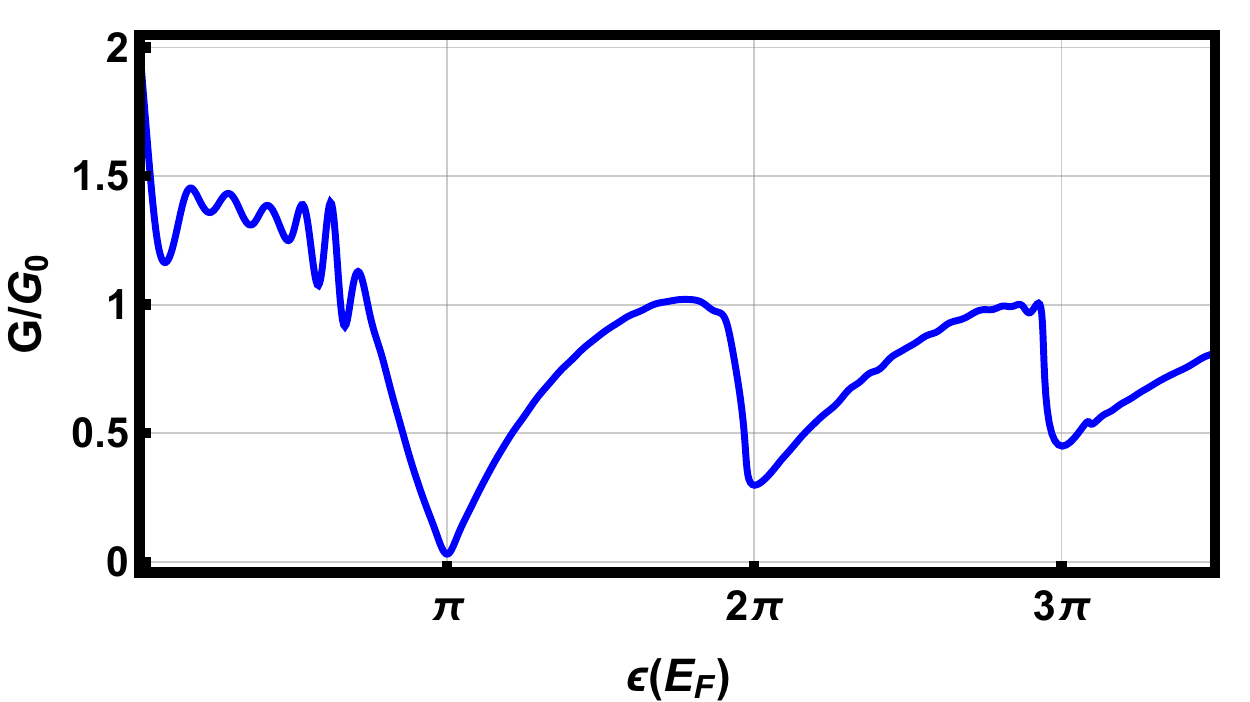}
        \label{FigG:SubFigD}
    }
    \caption{
		(Color online) Conductance for SSLGSL-3R versus incident energy $\varepsilon$ with
		$q_2=\frac{1}{3}$, $\mathbb{V}=1.5\pi$, 
		and four numbers of elementary cells $n=1,10,20,30$.
	}
	\label{FigG}
\end{figure}

Figure \ref{FigG} presents the conductance for SSLGSL-3R as function of incident 
energy $\varepsilon$ with $q_2=\frac{1}{3}$, $\mathbb{V}=1.5\pi$ and different values of 
elementary cells $n=1,10,20,30$. We observe that for one cell Figure \ref{FigG}\subref{FigG:SubFigA}, 
the conductance decreases from value $2$ of the pristine graphene to a value just below $1$ 
before the energy is equal to $\pi$ and starts after this value slightly oscillating. When $n$ 
increases the conductance shows minimums located near the levels of  VDPs at energies 
$k\pi, k\in\mathbb{N}^{*}$. In Figure \ref{FigG}\subref{FigG:SubFigB} ($n=10$), 
the transmission has oscillations between the ODP ($\varepsilon=0$) and the first  VDP ($\varepsilon=\pi$), 
the first conductance minimum is exactly at the first VDP but the other minimums are close to the relative  VDPs.
As long as $n$ increases (\ref{FigG}\subref{FigG:SubFigC}, \ref{FigG}\subref{FigG:SubFigD}),
the conductance minimums are placed exactly at the  VDPs levels and the oscillations decrease. 
We find that for $n\geq 30$, the conductance takes a stable form and therefore we realize it is one
corresponding 
to graphene superlattice. After the first  VDP, the conductance varies between $1$ and
the predict minimums.

\begin{figure}[!ht]\centering
    \subfloat[$n=1$]{
        \includegraphics[width=7cm]{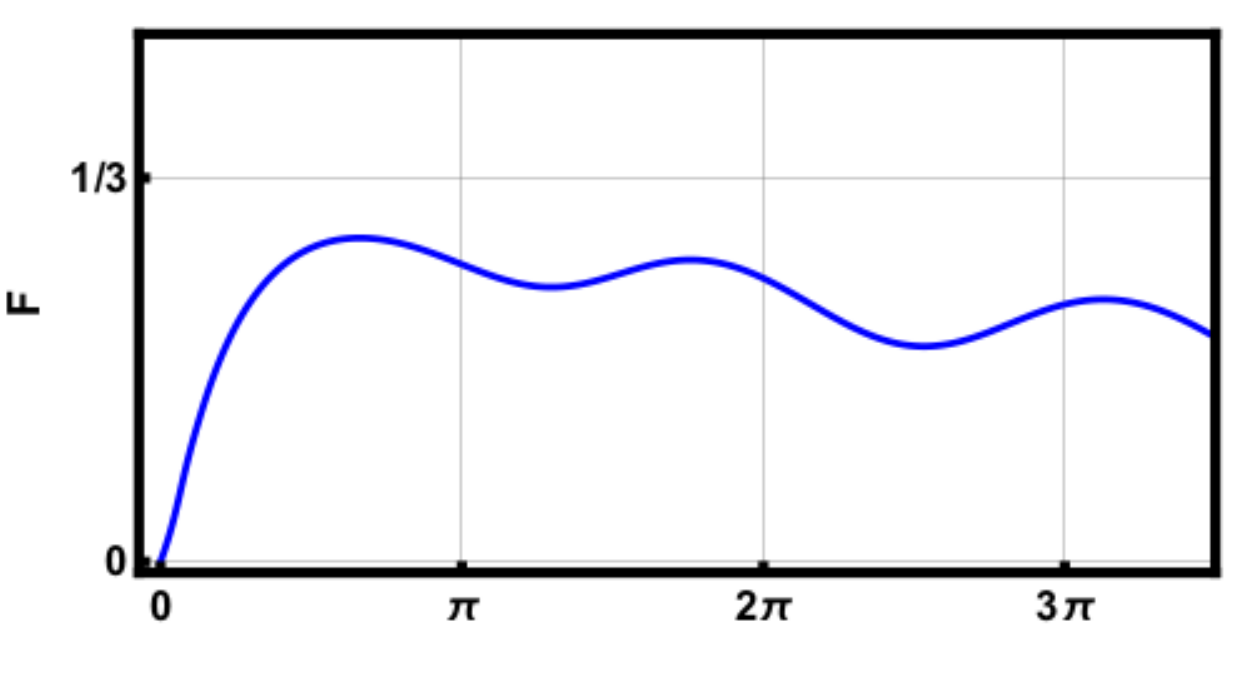}
        \label{FigFN:SubFigA}
    }\hspace{-0.3cm}
    \subfloat[$n=10$]{
        \includegraphics[width=7cm]{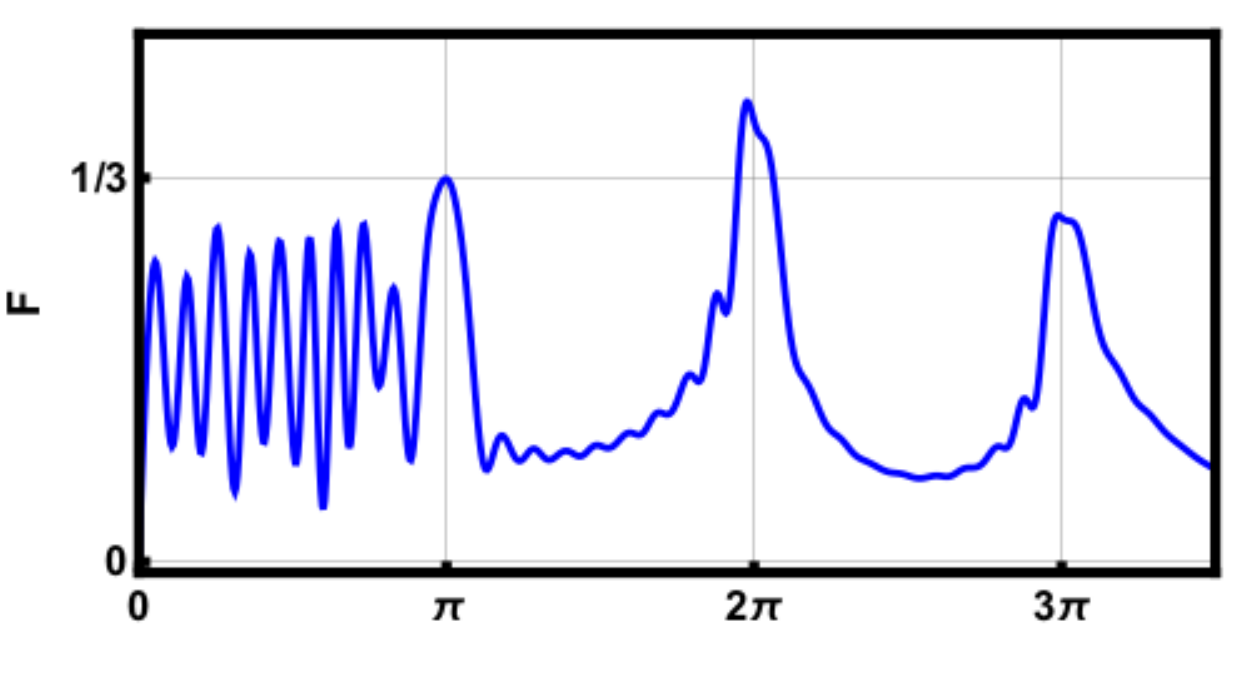}
        \label{FigFN:SubFigB}
    }\vspace{-0.5cm}
    \subfloat[$n=20$]{
        \includegraphics[width=7cm]{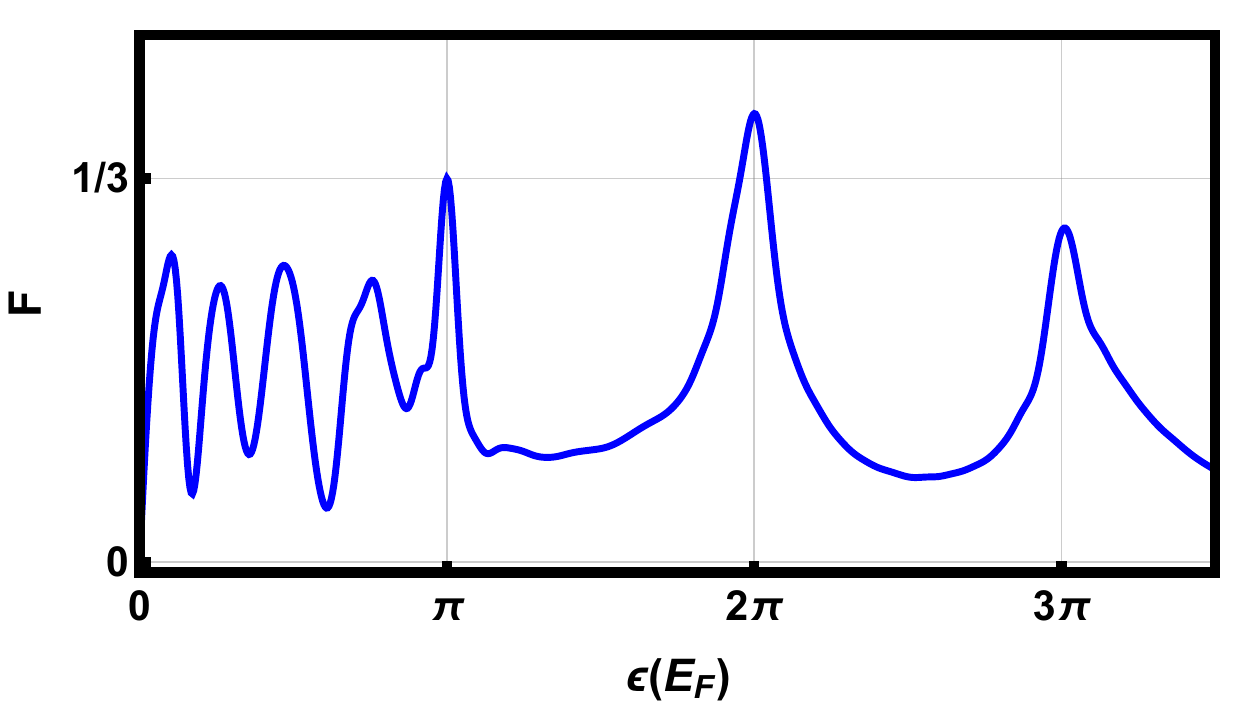}
        \label{FigFN:SubFigC}
    }\hspace{-0.3cm}
    \subfloat[$n=30$]{
        \includegraphics[width=7cm]{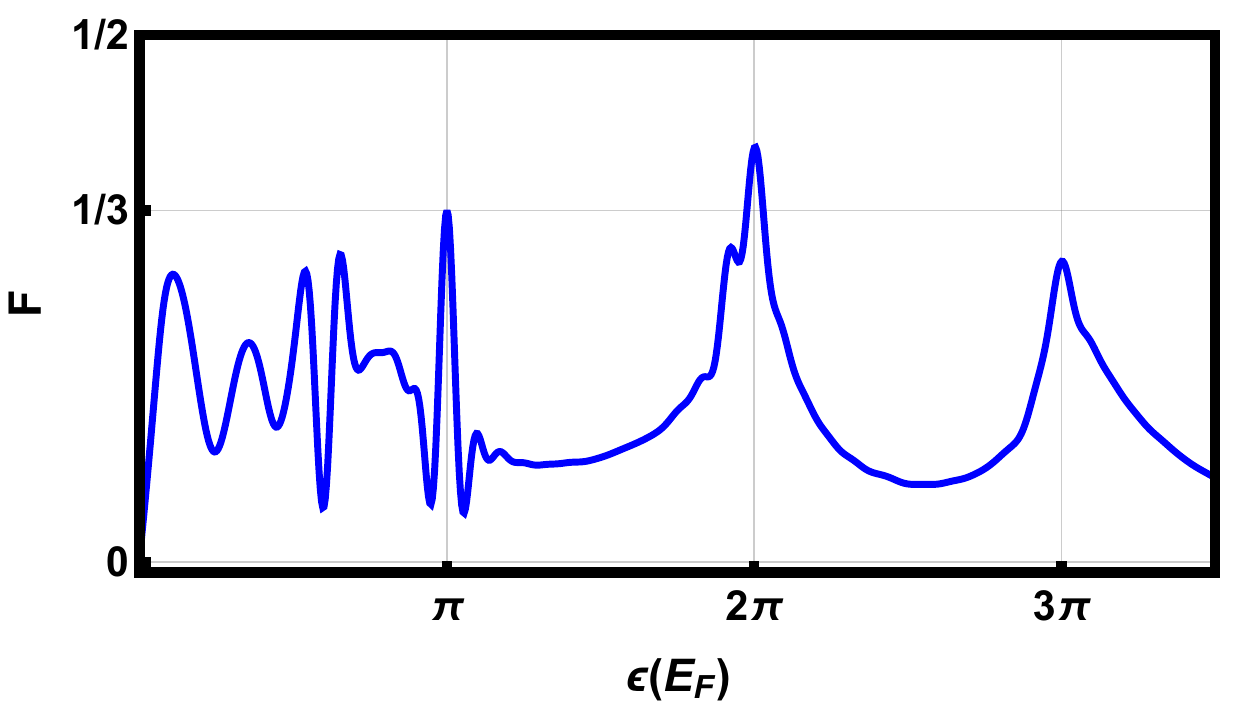}
        \label{FigFN:SubFigD}
    }
    \caption{
		(Color online) Fano factor for SSLGSL-3R versus incident energy $\varepsilon$ with
		$q_2=\frac{1}{3}$, $\mathbb{V}=1.5\pi$, $n=1,10,20,30$.
	}
	\label{FigFN}
\end{figure}

\begin{figure}[!ht]\centering
    \subfloat[$\varepsilon=\pi$]{
        \includegraphics[width=7cm]{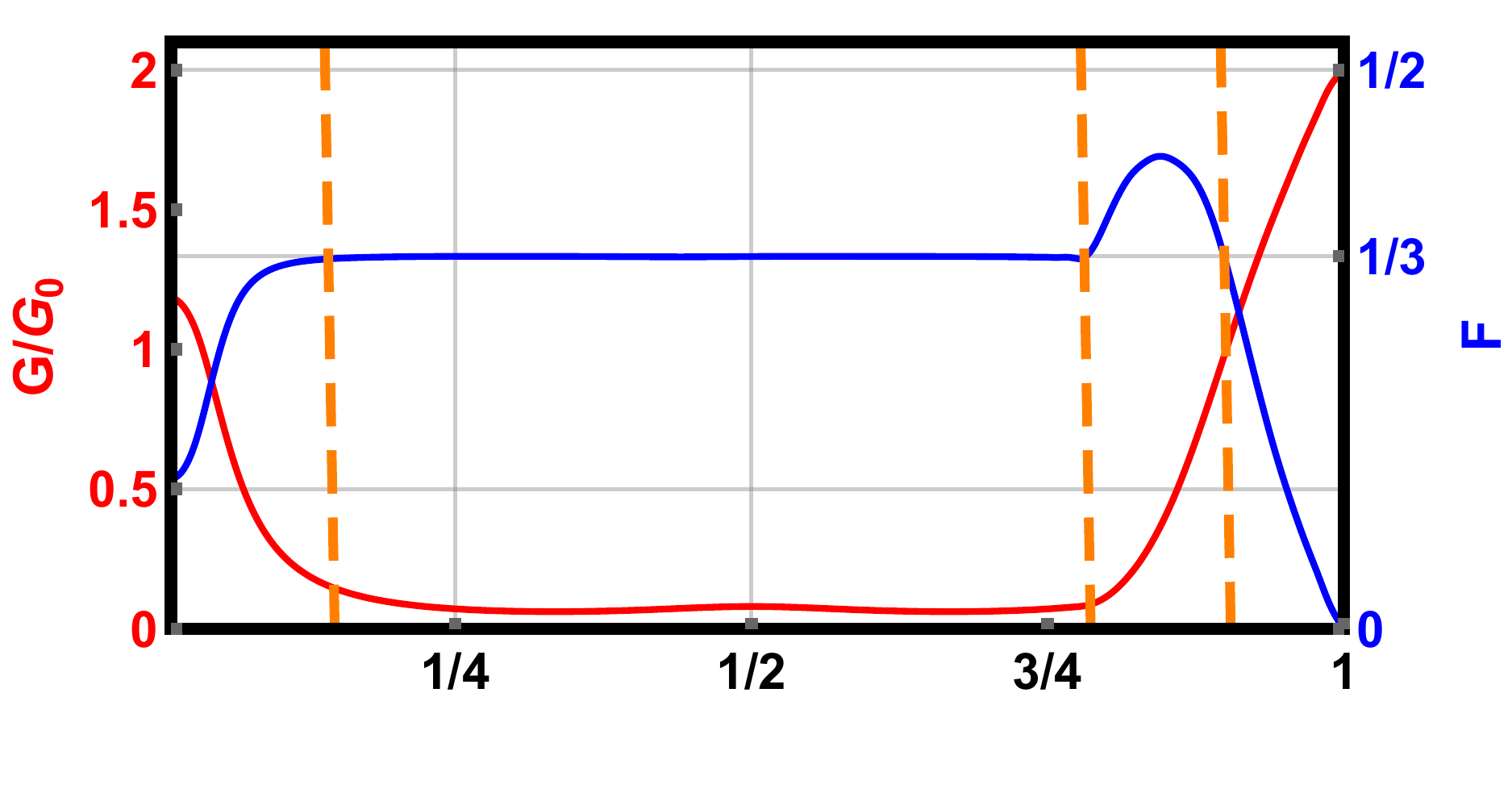}
        \label{FigGq:SubFigA}
    }\hspace{-0.1cm}
    \subfloat[$\varepsilon=2\pi$]{
        \includegraphics[width=7cm]{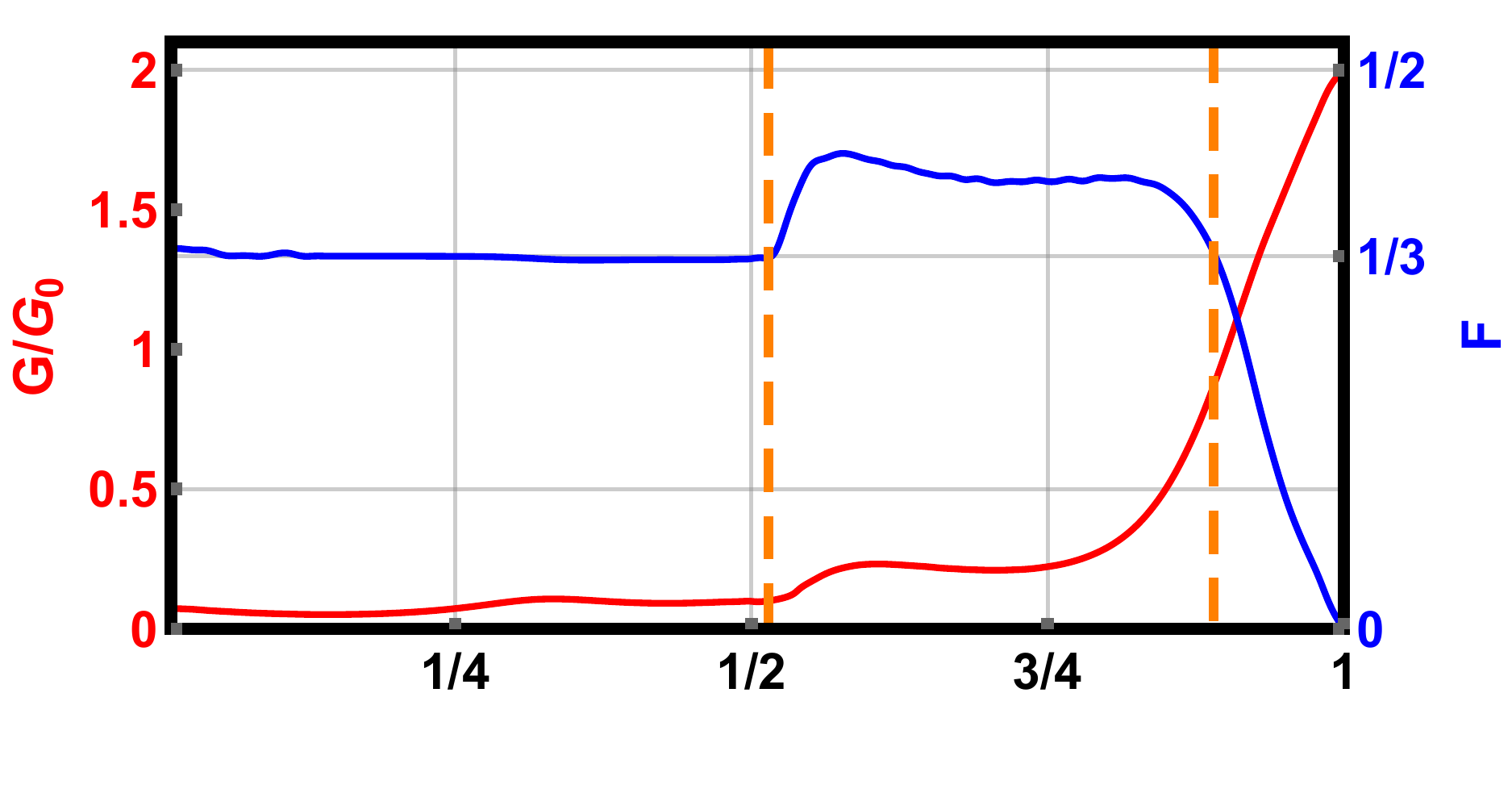}
        \label{FigGq:SubFigB}
    }\vspace{-0.5cm}
    \subfloat[$\varepsilon=3\pi$]{
        \includegraphics[width=7cm]{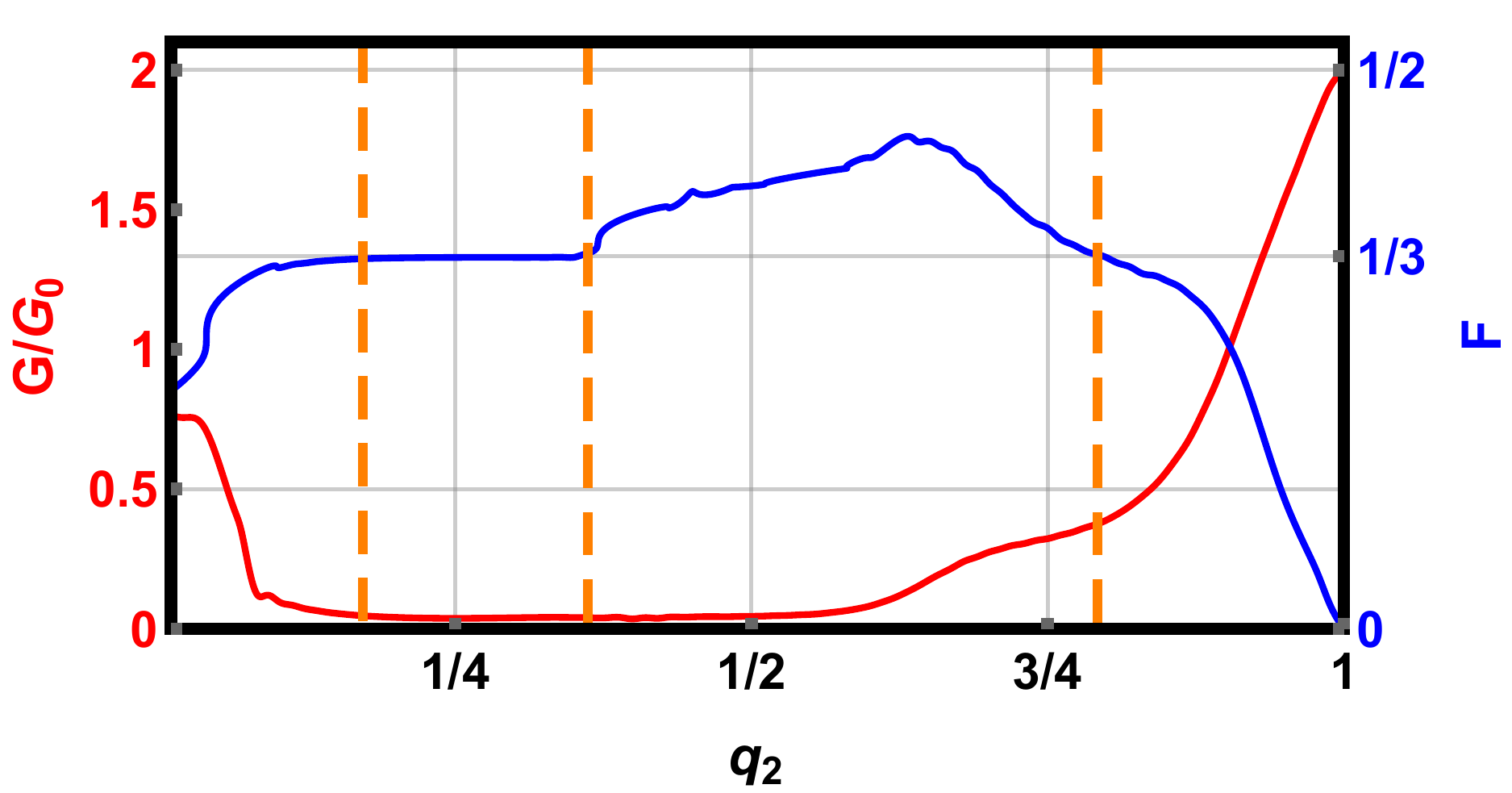}
        \label{FigGq:SubFigC}
    }\hspace{-0.1cm}
     \subfloat[$\varepsilon=4\pi$]{
         \includegraphics[width=7cm]{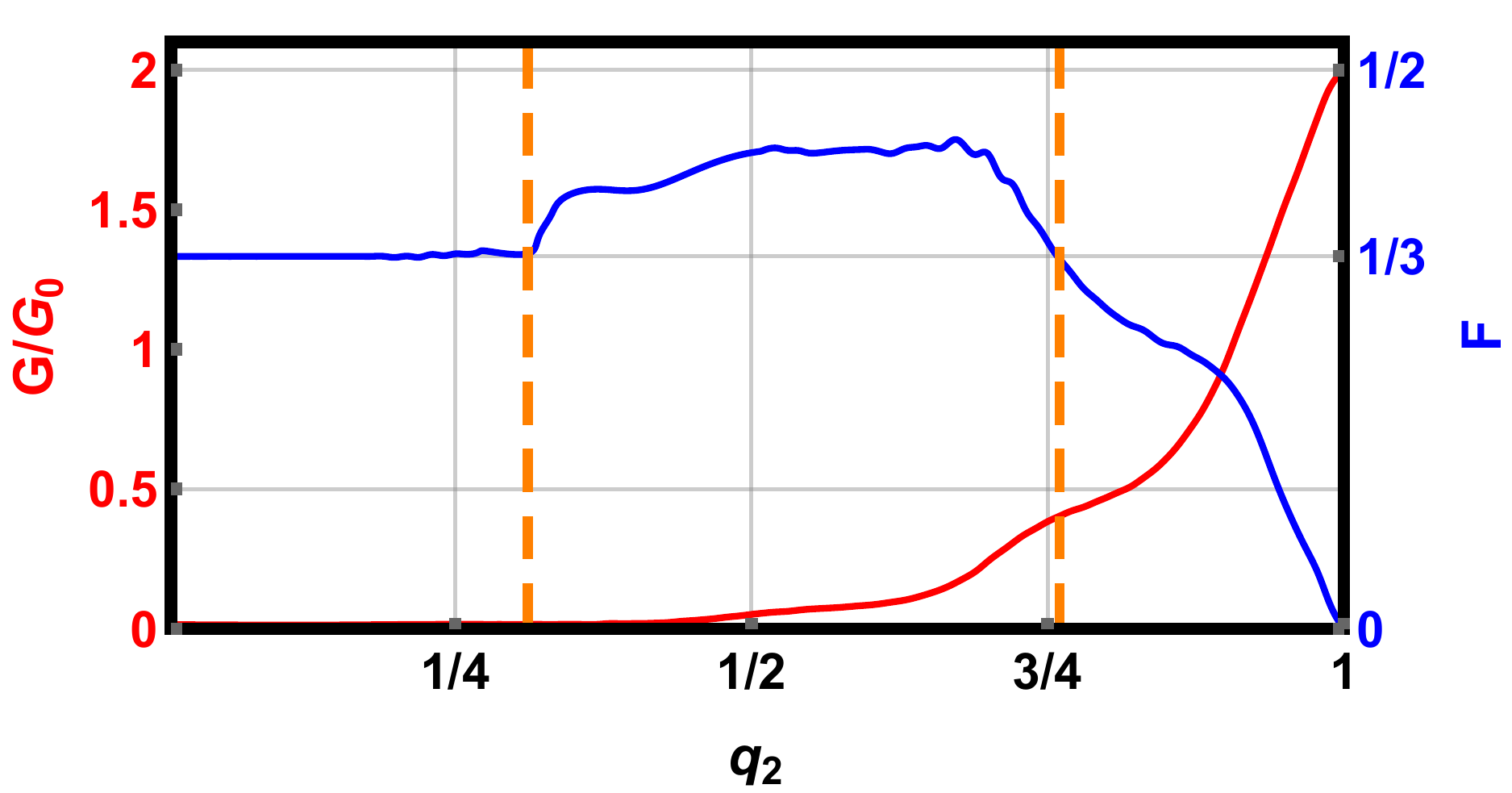}
         \label{FigGq:SubFigD}}
    \caption{
		(Color online) Conductance (red color) and Fano factor (blue color) for SSLGSL-3R 
		versus distance $q_2$
		for different energies $\varepsilon = \pi,2\pi,3\pi,4\pi$ with $\mathbb{V}=5\pi$
		and $n=30$.
	}
	\label{FigGE}
\end{figure}

Figure \ref{FigFN} shows the Fano factor versus incident energy $\varepsilon$
with $q_2=\frac{1}{3}$, $\mathbb{V}=1.5\pi$, for different values
$n=1,10,20,30$.
For one cell \ref{FigFN}\subref{FigFN:SubFigA}, the Fano factor increases from $0$ to value just lower than $1/3$ and starts to decrease again
to value higher than $0$. As long as the number $n$ increases, more than one maximum
of the Fano factor appear at $\varepsilon=m\pi$ with $m\in\mathbb{Z}$, as a result of the transmission gaps and
conductance. We observe that the Fano factor oscillates intensely at low energy for $n=10,~20,~30$ and as long as
$n$ increases, the amplitudes and frequencies of oscillation decrease. Note that, for the range of energy
$\varepsilon<\mathbb{V}$ the Fano factor does not exceed $1/3$. However, the Fano factor can exceed $1/3$
(for energy $\varepsilon= 2 \pi$), which is in agreement with the result found in \cite{XU2015188} for SSLGSL-2R.

Figure \ref{FigGE} shows the conductance $G/G_{0}$ (red line, red frame ticks  on left) 
and Fano factor $F$ (blue line, blue frame ticks   on right) as function of
 distance $q_2$ of intermediate region with  $\varepsilon = \pi,~2\pi,~3\pi,~4\pi$,
 $\mathbb{V}=5\pi$ and $n=30$.
 Let us see what happens in the extreme cases when two critical values of $q_2$ are inspected.
 Indeed, for  $ q_2 = 0 $ meaning that our system behaves as SLGSL-2R, we distinguish two interesting cases.
 Firstly when  $\varepsilon= \pi,~3\pi$ (odd values in $\pi$) according to Figures  
 \ref{FigGE}\subref{FigGq:SubFigA}, \ref{FigGE}\subref{FigGq:SubFigC}  
 $G/G_0$ is always in the interval $[1/2,3/2]$ and  $F$ is less than $1/3$.
 Secondly when
$ \varepsilon= 2\pi,~4\pi$ (even values in $\pi$) Figures 
\ref{FigGE}\subref{FigGq:SubFigB}, \ref{FigGE}\subref{FigGq:SubFigD} show that 
$G/G_{0}$ is almost null and $F$ is exactly equal $1/3$.
For $ q_2 = 1 $  our system is now a pristine graphene, we observe that for all incident energies
$ G/G_{0} = 2 $ and $F=0$, which are quit normal because 
in such situation the transmission is total $T_{30}=1$.
%
%
Still now the cases where $q_2$ is in the range $0< q_2<1 $, which means
our system is actually a  SLGSL-3R.
We can divide 
Figures \ref{FigGE}\subref{FigGq:SubFigA} and \ref{FigGE}\subref{FigGq:SubFigC}, 
in four zones according to the range taken by $q_2$, which are 
separated by orange dashed vertical lines. We observe that
%
$ G/G_{0}<3/2 $ and $F<1/3$ in first zone, 
$ G/G_{0}\rightarrow 0 $ and $F=1/3$ in second zone, 
$ G/G_{0}<3/2 $ and $F>1/3$ in third zone, $ G/G_{0}>3/2 $ and $F<1/3$ in forth. 
However
in Figures \ref{FigGE}\subref{FigGq:SubFigB} and \ref{FigGE}\subref{FigGq:SubFigD}, 
the first zone is omitted because $\mathbb{V}=5\pi$ does not coincide with the transmission 
bands intercalated alternately by gaps seeing in Figure \ref{FigTq2}\subref{FigTq2:SubFigB}.

\begin{figure}[!ht]\centering
     \subfloat[]{
         \includegraphics[width=7.5cm]{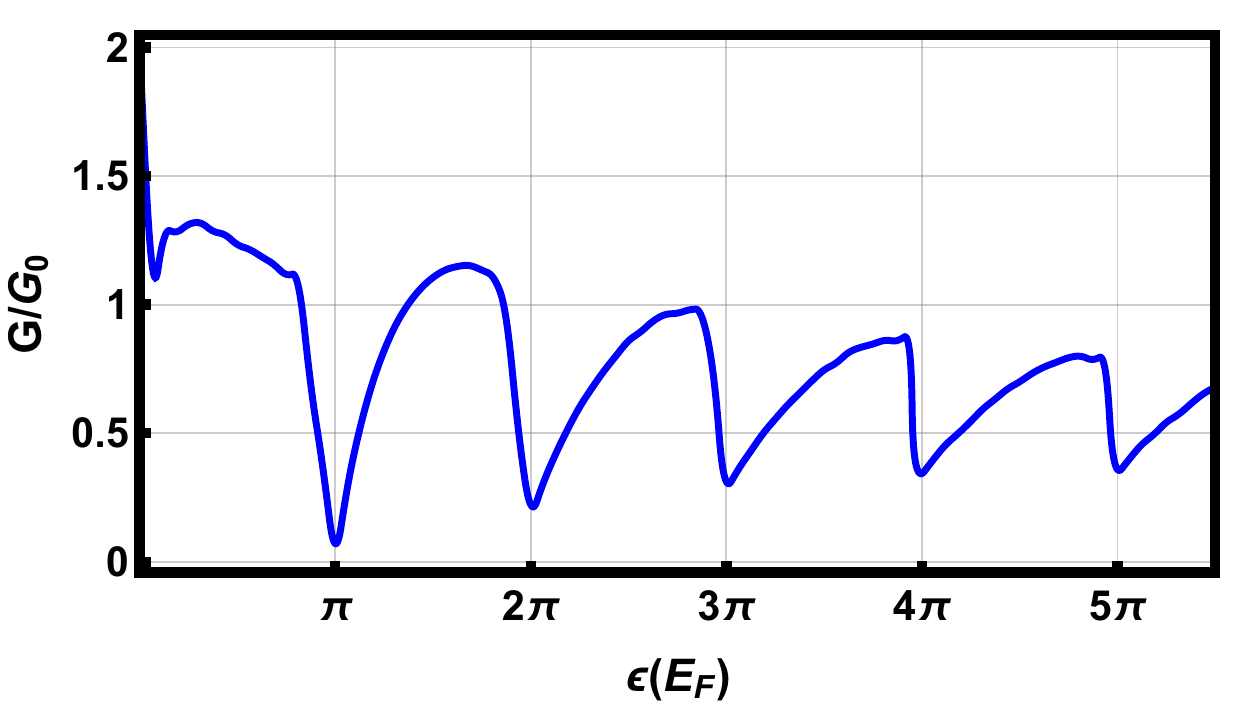}
         \label{FigGq2:SubFigA}
     }\hspace{-0.3cm}
    \subfloat[]{
        \includegraphics[width=7.5cm]{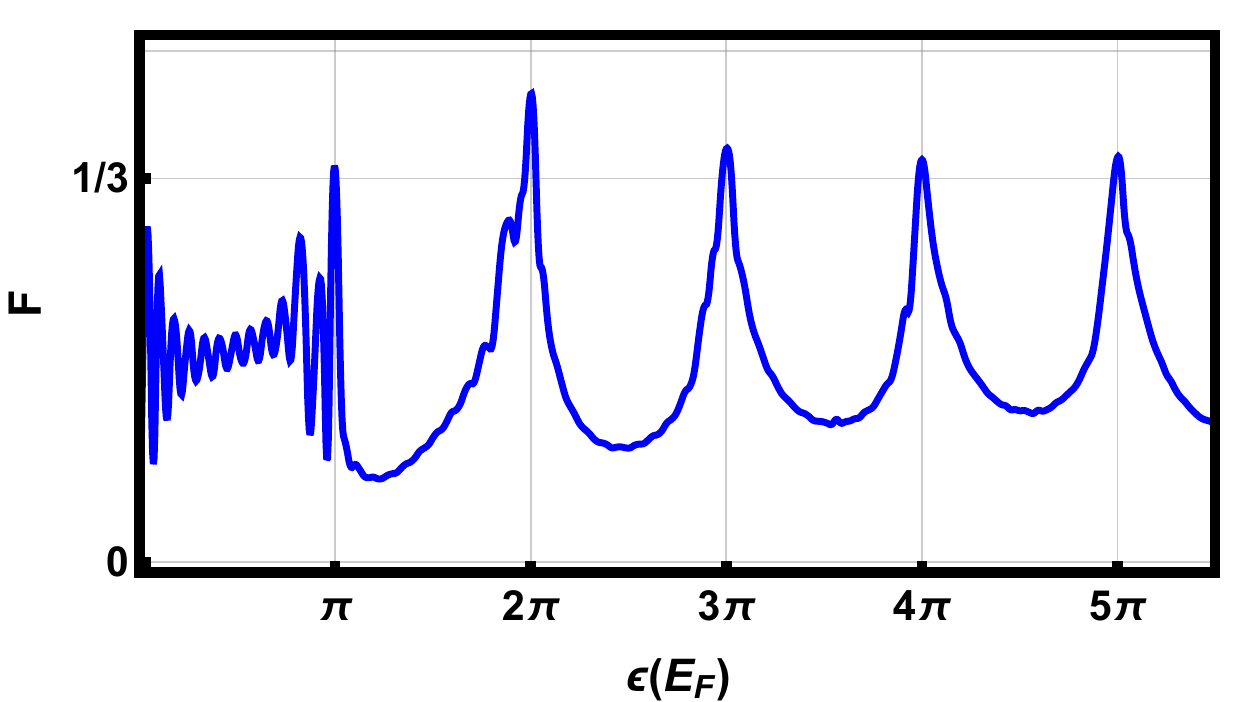}
        \label{FigFq2:SubFigB}
    }
    \caption{
		(Color online) 
		Conductance \protect \subref{FigGq2:SubFigA} and Fano factor 
		\protect\subref{FigFq2:SubFigB} for SSLGSL-3R versus incident energy $\varepsilon$
		with
		$\mathbb{V}=5\pi$, $q_2=\dfrac{3}{4}$.
	}
	\label{FigFq2}
\end{figure}

Let us show   the conductance  and Fano factor for SSLGSL-3R  
versus incident energy $\varepsilon$ by choosing the case   $ q_2 = 3/4$ in 
Figure \ref{FigFq2}.
%
We clearly observe that the minimum 
in the conductance $G/G_0$ at the vertical Dirac points is associated with 
a maximum in the Fano factor $F$. This result is in agreement  with
those obtained in literature  \cite{Danneau,Trauzettel2006,Jellal2012 }.


\section{Conclusion}\label{Sec:Conclusion}

Using the transfer matrix method and the Chebyshev polynomials of the second kind, we have investigated
the transmission probability, ballistic conductance and Fano factor of electrons tunneling through the
symmetrical single layer graphene superlattice with three regions (SSLGSL-3R). The corresponding
elementary cell is composed of three successive regions of potential ($\mathbb{V}$, 0, $-\mathbb{V}$),
and characterized by three parameters ($\mathbb{V}$, $d$, $q_2$) where $d$ is the elementary cell width
and $q_2$ is the width of second region. Explicit calculations showed that the three quantities are 
functions  the physical parameters characterizing our system, which allow us to make derive interesting
results and make different discussions.

Interesting numerical results concerning the SSLGSL-3R transmission probability, conductance and Fano
factor have been reported. It was shown that the transmission probability density plot has the same
behavior of electronic band structure for SSLGSL-3R with $k_x\in[0,\frac{\pi}{d}]$. More than one
transmission gap exists in the periodic potential structure. More transmission gaps can be obtained
by increasing the number of elementary cells. The transmission gaps appear at $\varepsilon=m\pi$ with
$m\in\mathbb{Z}$, which is the position of VDPs. For energies coinciding with band gaps in electronic
band structure, Dirac fermions have zero transmission except for the first near the ODP where
the transmission is zero up to the value $\varepsilon=k_y d$. The part of the band structure,
between zero and $\varepsilon=k_y d$, corresponds to the bound states of Dirac fermions where
transmission is zero. For potential $\mathbb{V}=m\pi$, in the interval
$0\leqslant \varepsilon\leqslant\mathbb{V}$, there is  $m$ transmission gap,
which is exactly on side of the $m$ VDPs. 

By setting $k_y=\varepsilon\sin\theta/d$, we show that when $\theta$ increases
the width of each transmission gap increases, the position of the center of each
transmission gap is the position of the VDPs, and the adjacent transmission gaps
are separated by $\Delta\varepsilon=\pi$. For the energies of the band structures
which are $\varepsilon\leqslant\pi$, the superlattice behaves like a more refractive
medium than that of the pristine graphene. For the energies
$\pi\leqslant\varepsilon\leqslant2\pi$, the reflection is total from a critical
angle corresponding to the boundary between the energy band and the first gap.
Beyond the second VDP, we have emergence of other gaps at the level of each VDP
with emergence of other angles that separate energy bands and gaps. We notice that
the number of these angles is $2m-1$ angles between the $m$ and $m+1$ VDPs.

Finally, the conductance and the Fano factor of the periodic potential
structure with different values of number of elementary cells are also studied.
It was shown that the minimums of conductance and maximums in Fano factor are
located exactly at the positions of VDPs. At low energy, the conductance and
Fano factor oscillate intensively and when the number of elementary cell $n$
increases, the amplitudes and frequencies of oscillation decrease.
In our case for $n>30$ the conductance takes a stable form attributed to the SSLGSL-3R.
The physical parameter $q_2$ is important to improve and control the conductance
and the Fano Factor. We end up our study by choosing the $q_2=3/4$ in order to determine
the conductance and its Fano factor in SSLGSL-3R.

\section*{Acknowledgment}

The generous support provided by the Saudi Center for Theoretical Physics (SCTP)
is highly appreciated by all authors.


\section{Appendix}

Given that $\Omega$ is a $2\times 2$ matrix, it turns out that one can write
its $n^{th}$ power in an elegant form involving the Chebyshev polynomials.
Indeed, $\Omega^2$ can be expressed as
\begin{equation}\label{22}
	\Omega^{2}=\Tr(\Omega) \Omega- \det(\Omega)\mathbb{I}=\Tr(\Omega) \Omega- \mathbb{I}
\end{equation}
and generally for any integer $n$, we have
 \begin{equation}\label{2311}
	\Omega^{n}=A_n \Omega+B_n{\mathbb I}
\end{equation}
where $A_n$ and $B_n$ are two coefficients of expansion. Pushing our iteration to the next order
\begin{equation}\label{23}
	\Omega^{n+1}=A_{n+1} \Omega+B_{n+1}{\mathbb I}
\end{equation}
multiplying \eqref{2311} by $\Omega$ and using \eqref{22} to end up with
\begin{equation}\label{27}
	A_{n+1}=\Tr(\Omega)A_n-A_{n-1}, \qquad A_n=-B_{n+1}
\end{equation}
which yields the three-term recurrence relations for the coefficient $A_n$. Next,
we will establish a mathematical tools to interpret \eqref{27}.

Let us now recall some interesting tools which concern the Chebyshev polynomials of the first
kind  \cite{masonchebyshev}
\begin{equation}\label{2701}
	T_0(z)=1,\qquad T_1(z)=2z,\qquad T_{\eta+1}(z)=2 z T_{\eta}(z)-T_{\eta-1}(z),\qquad
	\eta\geqslant 1
\end{equation}
as well as the second kind are defined
\begin{equation}\label{2702}
	U_0(z)=1,\qquad U_1(z)=z,\qquad U_{\eta+1}(z)=2 z U_{\eta}(z)-U_{\eta-1}(z),\qquad
	\eta\geqslant 1
\end{equation}
Using the above tools to show that coefficients $A_n$ are Chebyshev polynomials. Indeed, first we
set the suitable variable $z$ from \eqref{2702}
\begin{equation}
	z=\frac{1}{2}\Tr(\Omega)=\cos(\vartheta)
\end{equation}
where $\vartheta$ is Bloch phase of the periodic system. Second to determine what kind of Chebyshev polynomials,
we calculate $ A_{0}(z)$ and $A_{1}(z)$, then comparing \eqref{2311} for $n=2$ and \eqref{22} to write
\begin{equation}
    A_2=\Tr(\Omega)=2z,\qquad B_2=-\det(\Omega)=-1
\end{equation}
From \eqref{22}, \eqref{2311} and \eqref{27}, we obtain

\begin{equation}
    A_1=1,\qquad A_0=-B_1=0.
\end{equation}
Combining all to write
\begin{equation}
    A_0(\vartheta)=U_{-1}(\vartheta)=0,
    \qquad A_1(\vartheta)=U_0(\vartheta)=1,
    \qquad A_2(\vartheta)=U_1(\vartheta)=2z
\end{equation}
showing that $A_n$ is the Chebyshev polynomial of second kind $A_n(\vartheta)=U_{n-1}(\vartheta)$.
From \eqref{27} and \eqref{2311} we find
\begin{equation}\label{T32}
	\Omega^{n}=U_{n-1}(\vartheta)\Omega-U_{n-2}(\vartheta)\mathbb{I}
\end{equation}
where $U_n(\vartheta)$ is fixed by
\begin{equation}\label{a6}
	U_n(\vartheta)=\frac{\sin((n+1)\vartheta)}{\sin\vartheta}.
\end{equation}


\end{document}